\documentclass[12pt,preprint]{aastex}
\shorttitle{The Ca {\footnotesize II} Triplet as Metallicity indicator}
\shortauthors{Carrera et al.}

\begin{document}
\newcommand{\ki}{[Fe/H]$_{KI03}$}
\newcommand{\myemail}{skywalker@galaxy.far.far.away}


\title{The Infrared Ca {\footnotesize II} triplet as metallicity indicator}

\author{R. Carrera and C. Gallart}
\affil{Instituto de Astrof\'{\i}sica de Canarias, Spain}
\email{rcarrera@iac.es}
\email{carme@iac.es}

\author{E. Pancino}
\affil{Osservatorio Astronomico di Bologna, Italy}

\and

\author{R. Zinn}
\affil{Department of Astronomy, Yale University, USA}

\begin{abstract}

From observations of almost 500 RGB stars in 29 Galactic open and
globular clusters, we have investigated the behaviour of the infrared
Ca {\footnotesize II} triplet (8498, 8542 and 8662 \AA) in the age
range 13$\leq$Age/Gyr$\leq$0.25 and the metallicity range $-2.2\leq$
[Fe/H] $\leq$+0.47. These are the widest ranges of ages and
metallicities in which the behaviour of the Ca {\footnotesize II}
triplet lines has been investigated in a homogeneous way. We report
the first empirical study of the variation of the Ca{\footnotesize II}
triplet lines strength, for given metallicities, with respect to
luminosity. We find that the sequence defined by each cluster in the
Luminosity-$\Sigma$Ca plane is not exactly linear. However, when only
stars in a small magnitude interval are observed, the sequences can be
considered as linear. We have studied the the Ca {\footnotesize II}
triplet lines on three metallicities scales. While a linear
correlation between the reduced equivalent width ($W'_V$ or $W'_I$)
{\it versus} metallicity is found in the \citet{cg97} and
\citet{ki03} scales, a second order term needs to be added when the
\citet{zw84} scale is adopted. We investigate the role of age from the
wide range of ages covered by our sample. We find that age has a weak
influence on the final relationship. Finally, the relationship derived
here is used to estimate the metallicities of three poorly studied open
clusters: Berkeley 39, Trumpler 5 and Collinder 110. For the latter, the
metallicity derived here is the first spectroscopic estimate
available.
\end{abstract}




\section{Introduction}

The main functions defining the star formation history of a complex
stellar system are the star formation rate, SFR($t$) and the chemical
enrichment law, $Z(t)$, both function of time. The SFR($t$) can be
derived in detail from deep color--magnitude diagrams. $Z(t)$ has been
traditionally constrained by the color distribution of RGB
stars. However, this method of deriving metallicities from photometry
is a very crude one because in the RGB there is a degeneracy between
age and metallicity. To break this degeneracy we may obtain
metallicities from another source and then derive the age from the
positions of stars in the color--magnitude diagram. Of course, the
best way to obtain stellar metallicities is high-resolution
spectroscopy, which also provides abundances of key chemical
elements. However, a lot of telescope time is necessary to measure a
suitable number of stars. The alternative is low-resolution
spectroscopy, which allows us to observe a large number of stars in a
reasonable time using modern multi-object spectrographs.  At low
resolution, the metallicity is obtained from a spectroscopic line
strength index. The Mg$_2$, Ca {\footnotesize II} H \& K and Ca {\footnotesize II} infrared
triplet lines, the Fe lines, etc., are the most widely used indexes
for obtaining stellar metallicities. Different indexes are adequate
for different types of stars. For example, Fe lines are useful for
stars at the base of the RGB or in the main sequence
turn-off. Observation of these stars, however, is only possible for
the closest systems and even those require 8 m-class telescopes and
long integration times. Thus, for external galaxies, the only stars
that can be observed with modern multi-object spectrographs and
reasonable amounts of telescope time are those near the tip of the
RGB. A good spectroscopic index to obtain metallicities for these
stars is the infrared Ca {\footnotesize II} triplet (CaT), whose lines
are the strongest features in the infrared spectra of red giant stars.

\citet{az88} demonstrated that in the integrated
spectra of Galactic globular clusters, the equivalent widths of CaT
lines are strongly correlated with metallicity. As the near-infrared
light of globular clusters, where the CaT lines are, is dominated by
the red giant contribution, this relation may be also true in these
stars individually. Subsequent studies focused on the analysis of
individual red giants in globular clusters
\citep[e.g.][]{adc91}. These studies demonstrated that the strength of
the CaT lines changes systematically with luminosity along the
RGB. Moveover, for a given luminosity, the strength of these lines is
correlated with the cluster metallicity. Many authors have obtained
empirical relationships between the combined equivalent width of the
CaT lines and cluster metallicity. A very comprehensive work in this
field was published by \citet{r97a}, based on 52 Galactic globular
clusters covering a metallicity range of $-2\leq$ [Fe/H]
$\leq-0.7$. They compared the resulting calibration in the
\citet{zw84} and \citet{cg97} metallicity scales. While in the
\citet{cg97} scale a linear correlation between metallicity and
equivalent width of the CaT lines at the level of the
horizontal-branch (HB) V-V$_{HB}$=0 (known as reduced equivalent
width) was found for all clusters, this relationship was not linear
when the \citet{zw84} scale was used. In most studies, the run of CaT
lines with metallicity has been investigated in globular clusters
only, which have all similar ages. If we wish to derive stellar
metallicities in systems in which star formation has taken place in
the last few Gyr, such as dwarf irregular galaxies or open clusters,
it is necessary to address the role of age on the CaT strength. Some
authors have used (a few) young open clusters to study the behaviour
of the CaT with metallicity \citep[e.g.][]{sunt92}, using the
\citet{zw84} metallicity scale as reference. \citet{c04} very recently
obtained a new relationship, using open and globular clusters covering
$-2\leq$ [Fe/H] $\leq-0.2$ and 2.5 $\leq$ (age/Gyr) $\leq$ 13 in the
\citet{cg97} scale. They found a linear correlation among the reduced
equivalent width and metallicity. This indicates a weak influence of
age in the range of ages investigated (age $\geq$ 2.5 Gyr). However,
to apply this relationship to systems with star formation over the
last Gyr and/or with stars more metal-rich than the solar metallicity,
it is necessary to investigate its behaviour further for younger ages
and higher metallicities.

The purpose of this paper is to obtain a new relationship between the
equivalent width of the CaT lines and metallicity, covering a range as
wide as possible of age and metallicity. Our sample covers $-2.2\leq$
[Fe/H] $\leq$+0.47 and 0.25 $\leq$ Age/Gyr $\leq$ 13. The influence of
age and the variation of the CaT lines along the RGB are
investigated. In Section \ref{sample}, we present the cluster
sample. In Section \ref{obsdata}, the observations and data reduction
are described. The way in which the equivalent width of the the CaT
lines has been computed is described in Section \ref{catriplet}, where
the behaviour of the CaT with luminosity is also investigated. In
Section \ref{catmetallicityscale} we obtain the relationship between
the equivalent width of the CaT lines and metallicity, and we discuss
the influence of age and the [Ca/Fe] ratio in them. Finally, the
derived relationships are used in Section \ref{derivedmetallicities}
to obtain the metallicities of the open clusters Berkeley 39, Trumpler
5 and Collinder 110.

\section{Clusters Sample\label{sample}} 

\placetable{cluster sample}

To study the behaviour of the CaT lines with metallicity, we have
observed individual stars, with available V magnitudes, in 29 stellar
clusters (15 open and 14 globular). Of the 29 clusters in this sample,
27 also have I magnitudes available. This sample covers the widest
range of ages (0.25 $\leq$ Age/Gyr $\leq$1 3) and metallicities (2.2
$\leq$ [Fe/H] $\leq$ +0.47) in which the CaT lines have been observed
in a homogeneous way. The main parameters of the observed clusters are
listed in Table \ref{clustersample}. Our sample covers most of the
open clusters visible from the northern hemisphere with enough stars
above the red clump to get a good sampling of the RGB, and with
magnitudes easily reachable with the INT, WHT and 2.2 m CAHA
telescopes. In particular, the sample contains NGC 6705 (M11), a very
young open cluster (0.25 Gyr) with a well populated RGB, and NGC 6791,
one of the oldest open clusters ($\sim$9 Gyr), which is among the most
metal-rich clusters in our Galaxy ([Fe/H] $\sim$ +0.47). From the
south, using the VLT\footnote{Based on observations made with ESO
telescopes at Paranal observatories under programme 074.B-0446(B).}
and CTIO 4 m telescope, we observed four globular clusters, including
NGC 5927 and NGC 6528, which are among the most metal-rich globular
clusters in our Galaxy. The sample also includes the observations of 9
globular and 3 open clusters available at the ESO archive, whose
observations were carried out with the same instrumental
configurations as our own. With the purpose of investigating the
behaviour of the CaT lines with luminosity, we have observed stars
along the RGB in 5 clusters spanning our whole range of metallicities.

Table \ref{clustersample} presents a list of all the clusters in our
sample, together with their main characteristics: age, distance
modulus, reddening, reference metallicities in 3 scales (see Section
\label{catmetallicityscale}) and [Ca/H]. In total, 26 of the 29
observed clusters have metallicities in at least one of the three
scales. For the other 3 clusters (Collinder 110, Trumpler 5 and
Berkeley 39), we calculate their metallicities with the relationships
obtained here.

\placetable{starsample}
    
\section{Observations and Data Reduction\label{obsdata}}

\begin{figure}
\epsscale{1}
\plotone{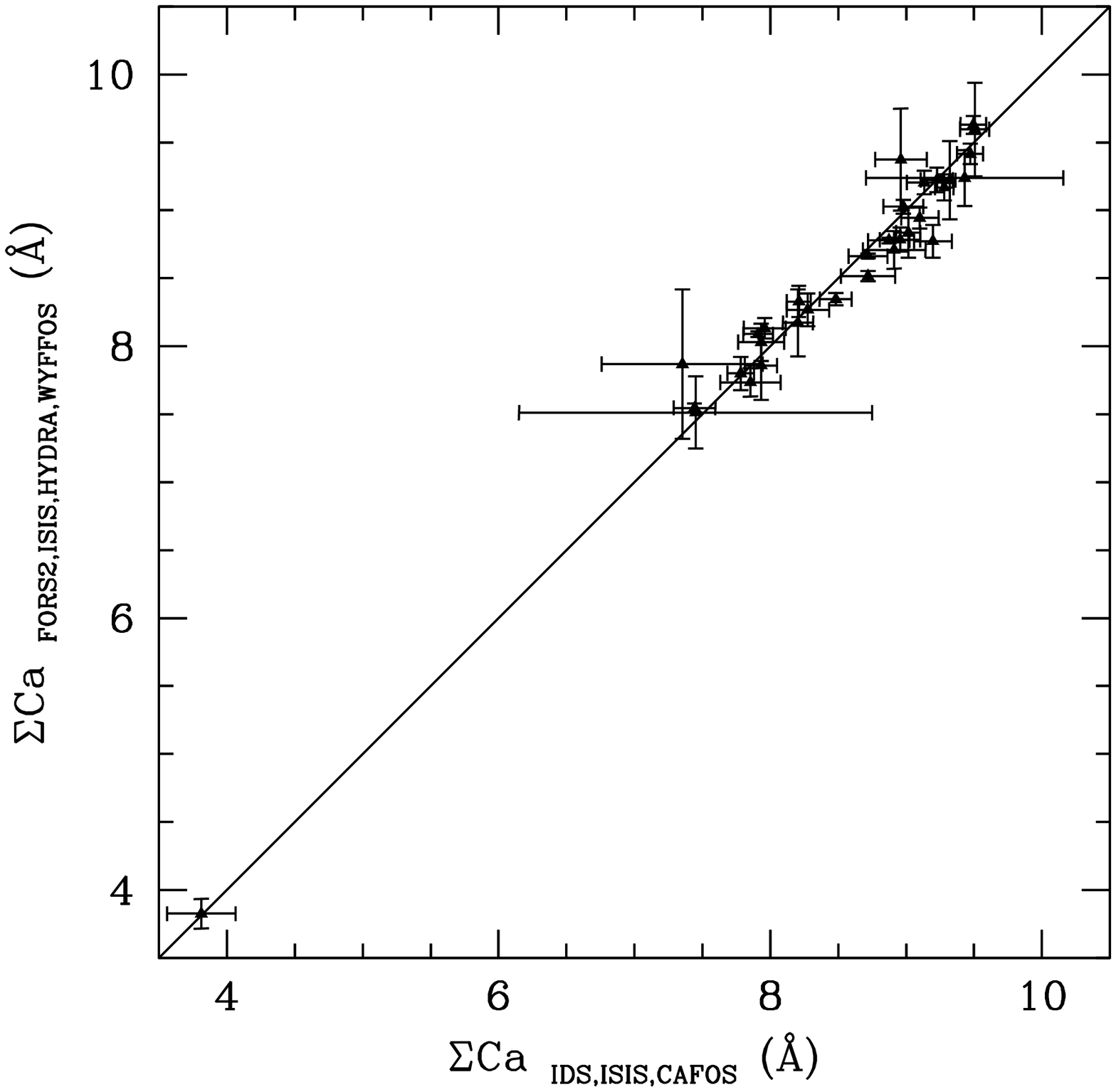}
\caption{Comparison between equivalent widths for stars observed with different telescopes. Small differences are within the
uncertainties.\label{telescopes}}
\end{figure}

About 500 stars have been observed in the 29 clusters of our sample in
6 different runs from 2002 to 2005, using the William Herschel
Telescope (WHT) and Isaac Newton Telescope (INT), both at Roque de los
Muchachos Observatory (La Palma, Spain), the 4 m telescope at CTIO (La
Serena, Chile), the 2.2 m at the Calar Alto Observatory (Almeria,
Spain) and the VLT at Paranal Observatory (Chile). The dates,
instruments and spectral resolution for each run are listed in Table
\ref{runs}. The instrumental configurations have been chosen in order
to ensure that the resolution was similar in each run.  The exposure
times were selected as a function of the magnitude of the stars in
order to obtain a good S/N, which in most cases was greater than
20. We have rejected from the analysis those stars with S/N lower than
20 (see below). In each run we have observed a few stars in common
with other runs in order to ensure the homogeneity of our
sample. Equivalent widths obtained for each star observed in two or
more runs have been plotted in Figure \ref{telescopes}. The
differences between runs are $< 0.1 \pm0.1$ \AA. The calculated
equivalent widths, together with the obtained radial velocity and the
utilized V and I magnitudes, are listed in Table \ref{starsample}.

The data taken with slit spectrographs, i.e., all except the observations
with HYDRA@CTIO and WYFFOS@WHT, were reduced following the procedure
described by \citet{massey92} using the IRAF\footnote{IRAF is
distributed by the National Optical Astronomy Observatory, which is
operated by the Association of Universities for Research in Astronomy,
Inc., under cooperative agreement with the National Science
Foundation.} packages but with some small differences described by
\citet{pont04}.  We obtained two images of each object, with the star
shifted along the slit. First, we subtracted the bias and overscan,
and corrected by the flat-field. Then, since the star is in a
different physical position in the two images, we subtracted one from
the other, obtaining a positive and a negative spectrum in the same
image.  With this procedure the sky is subtracted in the same physical
pixel in which the star was observed, thus minimizing the effects of
pixel to pixel sensitivity variations. Of course, a time dependency
remains since the two spectra have not been taken
simultaneously. These sky residues are eliminated in the following
step, when the spectrum is extracted in the traditional way and the
remaining sky background is subtracted from the information on both
sides of the star aperture. As the next step, the spectrum is
wavelength calibrated.  We then again subtracted the negative from the
positive (so we added both spectra because one is negative) to obtain
the final spectrum. Finally, each spectrum was normalized by fitting a
polynomial, excluding the strongest lines in the wavelength range such
as those of the CaT. The order of the polynomial changes among runs in
order to eliminate the response of each instrument. The wavelength
calibration of the VLT data (both from the archive and from run 6)
might be less accurate than the rest because arcs are not taken at the
same time and with the same telescope pointing as the object. The
effects of this on the wavelength calibration has discussed by
\citet{gallart01}, and we evaluate them in Section
\ref{radialvelocities}. However, since we are not interested in
obtaining precise radial velocities, this problem will not have an
important impact on our project.

HYDRA@CTIO and WYFFOS@WHT are multifibre spectrographs. The data
obtained with HYDRA has been extracted with the DOHYDRA task within
IRAF in the way described by \citet{valdes92}.  This task was
developed specially to extract data acquired with this instrument. The
procedure is described in depth by \citet{carrera05}. Basically, after
bias, overscan subtraction and trimming, DOHYDRA traces the apertures,
makes the flat-field correction and calibrates in wavelength. We
followed a similar procedure with the data obtained with WYFFOS, but
in this case we used the general DOFIBERS task, which works similarly
to DOHYDRA. Although both tasks allow for sky subtraction, the results
were poor, and important residuals of sky lines remained. To remove
the contribution of these sky lines, we have developed our own
procedure to subtract them. Basically, it consists in obtaining an
average sky spectrum from all fibres placed on the sky in a given
configuration. Before subtracting this average, high S/N sky, from
each star spectrum, we need to know the relation between the intensity
of the sky in each fibre (which varies from fibre to fibre due to the
different fibre responses) and the average sky. This relation is a
weight (which may depend on wavelength) by which we must multiply the
average sky spectra before subtracting it from each star. To calculate
it, we have developed a task that finds the weight which minimizes the
sky line residuals over the whole spectral region considered. As a
result of this procedure, the sky emission lines are removed very
accurately. Finally, the normalization was carried out in the same way
as previously described.

\begin{figure}
\epsscale{1}
\plotone{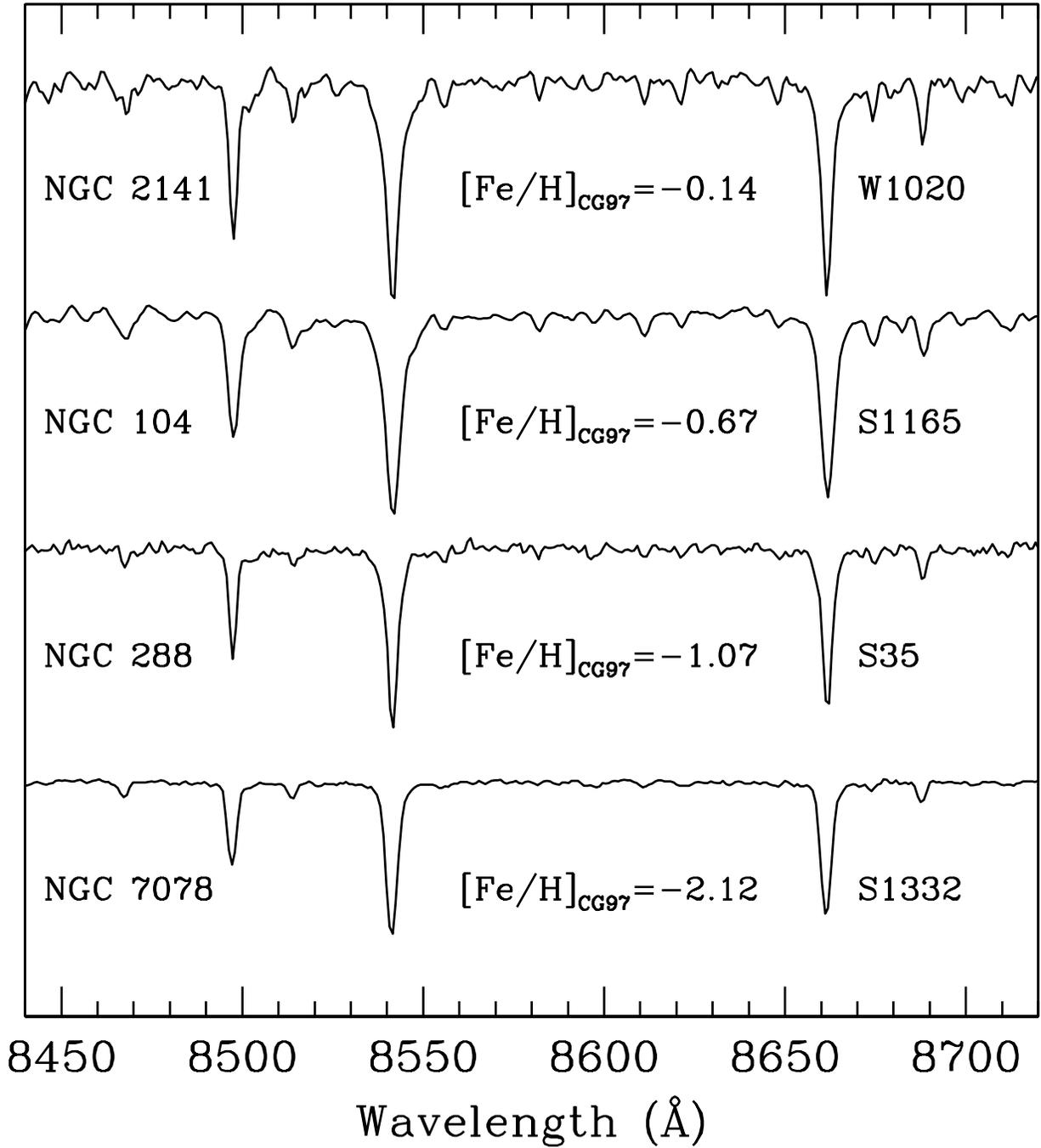}
\caption{Spectra of four stars in clusters with different metallicities. The metallicity decreases from top to bottom. Note
how the strength of the Ca {\footnotesize II} triplet lines increases with metallicity.\label{spectra}}
\end{figure}

Examples of 4 stars with different metallicities are shown in Figure
\ref{spectra}. Note how the strength of the CaT lines increases with
metallicity.

\placefigure{spectra}


The radial velocity of each star has been calculated in order to
reject cluster non-members. We used the FXCOR task in IRAF, which
performs the cross-correlation between the target and template spectra
of known radial velocity \citep{td79}. We selected between 8 and 10
template stars in each run that had very high S/N and covered a wide
range of radial velocities. The velocities were corrected to the
heliocentric reference frame within FXCOR. The final radial velocity
for each star was obtained as the average of the velocities obtained
from each template, weighted by the width of correlation peaks.

In the case of observations with slit spectrographers, the star might
not be exactly positioned in the centre of the slit. This error means
a velocity uncertainty given by $\Delta v=c\times \Delta \Theta \times
p/\lambda_0$, where: $c$ is the light speed, $p$ is the spectral
resolution given in \AA\ arcsec$^{-1}$; $\lambda_0$ is the wavelength
of the lines (in this case $\sim$8600 \AA), and $\Delta \Theta$ is the
angular offset of the star from the centre of the slit in arcsec. This
effect has been described by \citet{irwint02} and \citet{hz06}. In our
case, it may only be significant in the case of the VLT
observations. To estimate the offset in this case we used through-slit
images obtained at the beginning of the observation of each
configuration, taken to check that the stars were positioned in the
slits.  In this image we have measured the position of each stellar
centroid, which is compared with the position of the slit given in the
header of the image. The difference between both, $\Delta \Theta$,
allows us to calculate the uncertainty in the measurement of the
radial velocity. This value changes from one star to another, the
error being about 15 km s$^{-1}$ on average.

The mean velocity for each cluster is listed in Table
\ref{samplevr}. Most of the values obtained agree, within the
uncertainties, with previous measurements from the literature, even in
the case of the clusters observed with the VLT, where the
uncertainties are larger. In the case of NGC 2141, we found a mean
velocity similar to the value obtained by \citet{c04}. Both values
differ by 20 and 30 km s$^{-1}$, respectively, from the value found by
\citet{f02}. For Collinder 110, no previous measurement of its radial
velocity could be found in the literature.

\section{The Calcium Triplet \label{catriplet}}

We are interested in obtaining metallicities from red giant stars, and
within this group, from the brightest ones, which are of spectral
types K and M. The main features in the infrared spectra of these
stars are the CaT lines. But their spectra also contains other weak
atomic lines. The Fe {\footnotesize I} (8514.1, 8674.8, 8688.6 and
8824.2 \AA) and Ti {\footnotesize I} (8435.0 \AA) lines are the most
important. When within this range, we move to later spectral types,
and hence to cooler stars, molecular bands begin to appear that change
the slope of the local continuum.  The main contribution are from the
titanium oxide (TiO) bands, the strongest of which are the triplet
situated at 8432, 8442 and 8452 \AA~and the doublet at 8859.6 and
8868.5 \AA. There are other weaker bands at 8472, 8506, 8513, 8558 and
8569 \AA, near the bluest lines of the CaT. There are also several
vanadium oxide (VO) bands at 8521, 8538, 8574, 8597, 8605, 8624, 8649
and 8668 \AA. The strength of these features increases when the
temperature decreases, i.e.\ when we move to later spectral types. The
presence of these bands complicates the definition of the continuum,
which makes it difficult to obtain the equivalent widths of the CaT
lines for stars with T$_{eff}\leq$3500 K or (V-I)$>$2, in the most
metal-rich clusters. The description of the CaT region for other
spectral types can be found in \citet{cen01}.
   
\subsection{Definition of Line and Continuum Bandpass Windows}

\begin{figure}
\epsscale{1}
\plotone{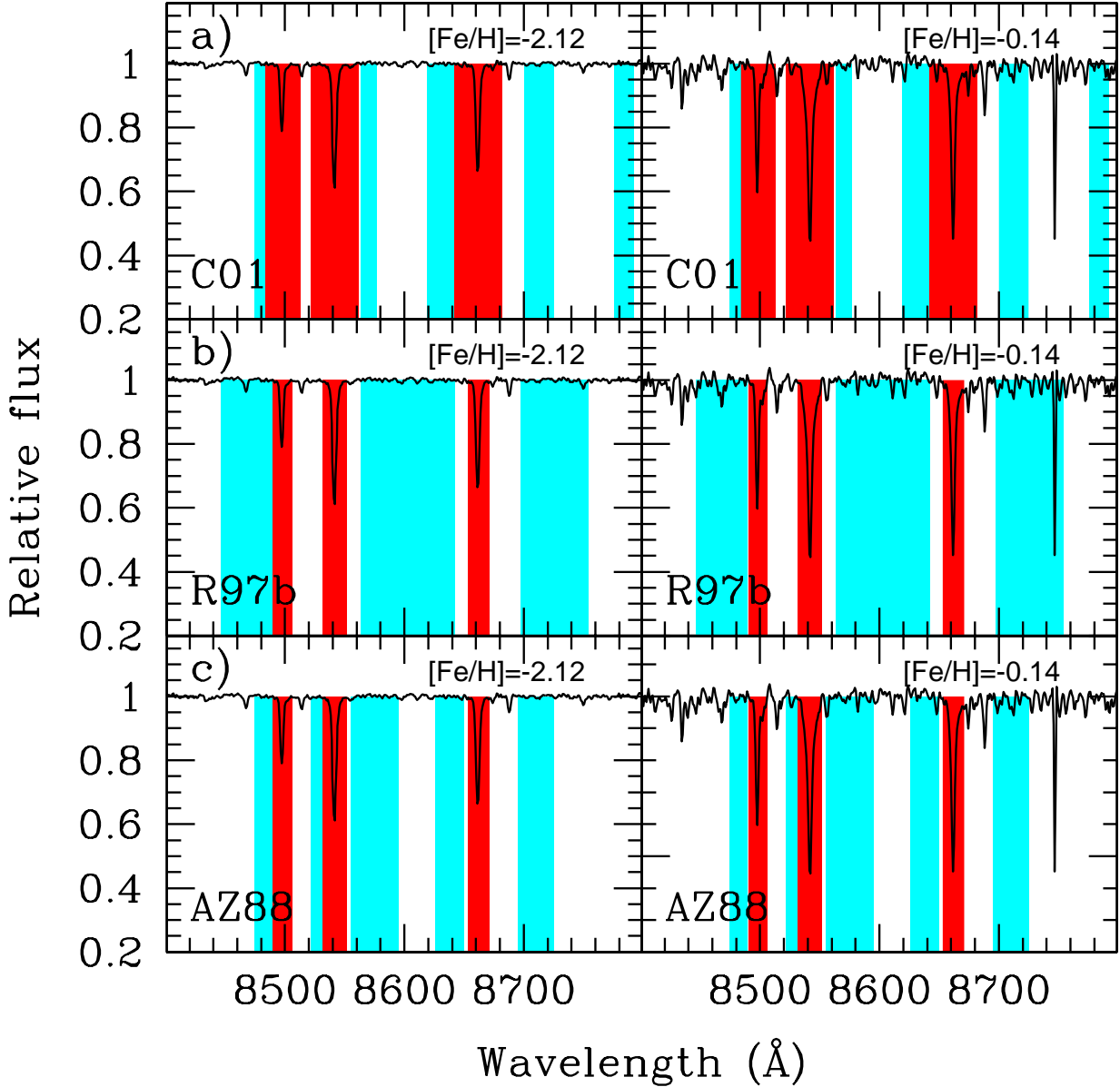}
\caption{Continuum (clear) and line (dark) bandpasses defined by (a) 
\citet{cen01}, (b) \citet{r97a} and (c) \citet{az88}. They
have been overplotted on to metal-poor (left) and metal-rich (right)
stars. The bands of \citet{cen01} are wider in the lines to cover the
wings fully and narrower in the continuum in order to avoid the most
prominent molecular features for metal-rich stars.\label{bandas}}
\end{figure}

In the literature we can find different prescriptions to measure the
strength of the CaT lines. The classical definition of a spectral
index consists in establishing a central bandpass covering a spectral
feature and one or more bandpasses on both sides to trace the local
continuum reference level. \citet{cen01} have presented a description
of the previous CaT index definitions and a comparison among them. In
Figure \ref{bandas} we have plotted the line and continuum bandpasses
used in several reference works, \citet{cen01} (a), \citet{r97a} (b)
and \citet{az88} (c), over a metal-poor (left) and a metal-rich
(right) spectrum. The \citet{az88} and \citet{r97a} indices were
defined for relatively metal-poor RGB stars where the influence of the
molecular bands is not important. The index of \citet{cen01} was
defined specifically to avoid the presence of molecular bands. Also,
from Figure \ref{bandas}, we can easily see that the wings of the
lines are larger than the line bandpasses defined by \citet{az88} and
\citet{r97a} in the case of the metal rich stars. Only the line
bandpasses defined by \citet{cen01} completely cover the line
wings. Although we have selected the bandpasses defined by
\citet{cen01}, which are listed in Table
\ref{bandastable}, the equivalent width of the line will be measured
in a different way, as described in the following section.
\placefigure{bandas}

\subsection{Equivalent widths\label{ew}}

The next step is to measure the line flux from its equivalent
width. The equivalent width of a spectral line can be measured in
different ways. One method is by numerical integration of the observed
spectra in a line band \citep[e.g.][]{cen01}. However, in the wings of
the strongest lines of the CaT there are some weak lines, whose
strength may change with different stellar atmospheric parameters
than the CaT lines. These lines must be excluded when we measure the
CaT equivalent width. The alternative \citep[e.g.][]{r97a,c04}
consists in fitting an empirical function to a line profile and
calculating the equivalent width from the integration of this fit.
Many functions have been used to fit the CaT line profiles, most
commonly a Gaussian profile \citep[e.g.][]{adc91}. However, as
\citet{c04} have shown, the Gaussian profile provides a good fit for
weak-line stars, but the fit is worse in strong-line stars, where the
contribution of the non-Gaussian wings of the CaT lines becomes
substantial. We have to take this point into account because the main
contributors to the strength of the CaT lines are their wings, while
the core is not very sensitive to the atmosphere and stellar
parameters \citep{erdelyi91}. \citet{r97a} fitted a Moffat function of
exponent 2.5. As \citet{pont04} has demonstrated, the behaviour of
Moffat function of exponent 2.5 is similar to the Gaussian fit for the
weakest lines. However, neither provides a good fit to the strongest
lines. \citet{c04} fitted the whole line profile with the sum of a
Gaussian and a Lorentzian function, which provides a better fit for
the strongest lines and agrees with the single Gaussian fit for the
weakest lines \citep[see][for a further discussion]{c04}. We have
compared the different functions in order to evaluate the quality of
the fit in the whole range of line strengths. We have chosen the sum
of a Gaussian and a Lorentzian function because this provides the best
fit for the whole range of equivalent widths in this study. We have
also checked whether a simple Gaussian or Moffat function would
produce a good fit in the case of spectra obtained with lower
resolution. Also in this case a Gaussian plus a Lorentzian provides
the best fit for strong-line stars.

A Gaussian plus a Lorentzian function has therefore been fitted to the
line profiles with a least-squares method, using the
Levenberg-Marquardt algorithm. For the whole range of equivalent
widths covered in this work, the differences between the observed line
and the fit are negligible for stars with S/N $\geq$ 20. Stars
with poorer S/N have been rejected. The equivalent width of each
line is the area limited by the fitted profile of the line and the
continuum level, defined as the linear fit to the mean values of the
flux in each window chosen to determine the continuum. Formal errors
of the fit are estimated as the difference between the equivalent
width measurement for continuum displacements of $\pm(S/N)^{-1}$.

\subsection{The CaT index \label{catindex}}

\begin{figure}
\epsscale{0.6}
\plotone{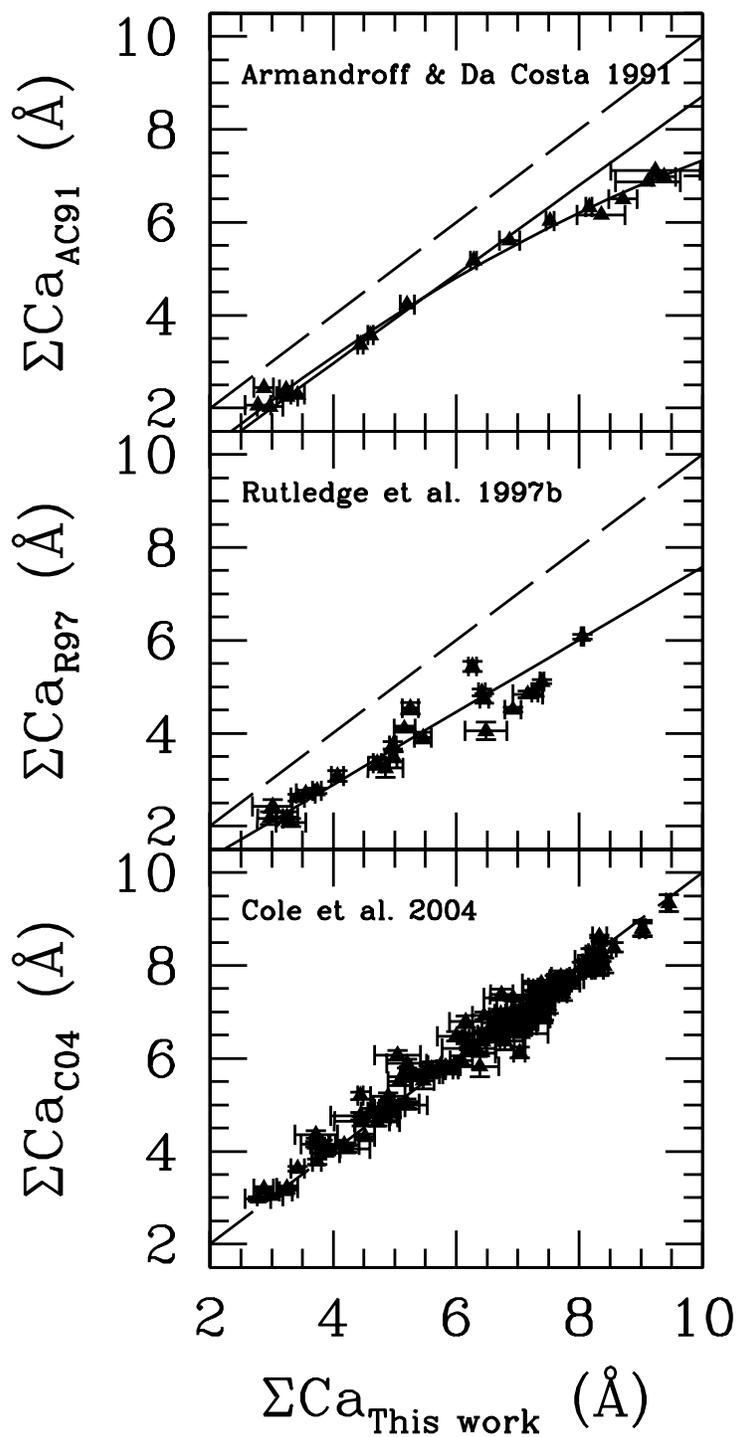}
\caption{Comparison between $\Sigma Ca$, as defined by \citet{adc91}, 
\citet{r97b} and \citet{c04}, and the values obtained in this paper.
 The dashed lines represent the one-to-one equivalence. Solid lines
 are best fits to the data.\label{comp}}
\end{figure}

The equivalent widths of the three CaT lines are combined to form the
global index $\Sigma$Ca \citep{adc91}. Some authors excluded the
weakest line at 8498 \AA\ on the basis of its poor S/N
\citep[e.g.][]{sunt93, c00}. Others have used all three lines, either
weighted \citep[e.g.][]{r97a} or unweighted \citep[e.g.][]{ol91}. As
our spectra have high S/N ratios, we used the unweighted sum of the
three lines, $\Sigma Ca$=W$_{8498}$+W$_{8542}$+W$_{8662}$, and
we calculate its error as the square root of the quadratic sum of the
errors of each line. As we have some stars in common with previous
works, we can compare the $\Sigma Ca$ calculated by us with values
obtained in previous papers. \citet{r97a} compared their $\Sigma$Ca
with previous index definitions until 1997. Here, for simplicity, we
are only going to compare our index with three reference works. Stars
in common with \citet{adc91, r97a} and \citet{c04} are plotted in
Figure \ref{comp}. As mentioned before, the works of \citet{adc91} and
\citet{r97a} were focused on old and metal-poor stars. However,
\citet{ol91} and \citet{sunt93}, using the same index as \citet{adc91}  
defined for globular cluster stars, measured the equivalent width of
the CaT lines in stars of two open clusters, M11 and M67,
respectively. We are going to use these values to complete the
measurements of \citet{adc91}.

\placefigure{comp}

We find a quasilinear relation up to $\Sigma Ca\sim$7 among the
$\Sigma Ca$ values in this paper and those obtained by
\citet{adc91} \citep[see also][]{sunt93}. From this point the relationship
saturates: while our index increases by an additional $\Delta\Sigma
Ca\sim$2, theirs only increases by $\Delta\Sigma Ca\sim$1.5 (on their
scale). We believe that the reason for this is that they fitted the
line profile by a Gaussian function which underestimates the
contribution of the line wings in strong lines (see Section~
\label{catindex}).  Note also the zero-point difference between both
scales. The relation is not exactly one to one because they did not
use the equivalent width of the weakest CaT line. However, the slope
close to one of the linear fit for the metal-poor stars implies that
the two indices are almost equivalent for these kind of stars. The
loss of linearity for strong-line stars partly explains why these
authors found a nonlinear relationship between the CaT index and
metallicity, but, of course, the metallicity scale also plays a role
in this issue, as we discuss in Section \ref{othermetallicities}. The
linear fit for $\Sigma$Ca$\leq$7 (solid straight line in top panel of
Figure
\ref{comp}) is:\\

\begin{equation}
\Sigma Ca_{AC91}=-0.88(\pm0.08)+0.96(\pm0.01)\Sigma Ca_{TP}
\end{equation}

and the second order polynomial fit for the whole range of equivalent widths is\\

\begin{equation}
\Sigma Ca_{AC91}=-1.10(\pm0.08)+1.20(\pm0.03)\Sigma Ca_{TP}-0.04(\pm0.01)\Sigma Ca_{TP}^2
\end{equation}

In the case of \citet{r97a}, who only observed stars with [Fe/H]
$\leq-0.7$, we find a linear correlation for the whole range of
equivalent widths. In this case the slope is less than one, meaning
that their index is less sensitive to changes in the strength of the
CaT lines than ours. For the same star, our index is higher than the
\citet{r97a} one. The linear fit is:\\

\begin{equation}
\Sigma Ca_{R97}=-0.23(\pm0.06)+0.78(\pm0.01)\Sigma Ca_{TP}.
\end{equation}

Finally, the correlation between \citet{c04} index and ours is one to
one ($\Sigma Ca_{TP}-\Sigma Ca_{C04}=0.009\pm0.0007$). As we used the
same empirical function and index definition of $\Sigma$Ca as
\citet{c04}, differences could only come from the definition of line
and continuum bandpasses. This means that, in the range of equivalent
widths covered here, both indices are equivalent. However, as the
continuum in our index has been defined to avoid the influence of TiO
bands, we expect that our index would also behave well in stars whose
continuum is contaminated by TiO bands.

\subsection{The reduced equivalent width\label{reducedew}}

\begin{figure}
\epsscale{1}
\plotone{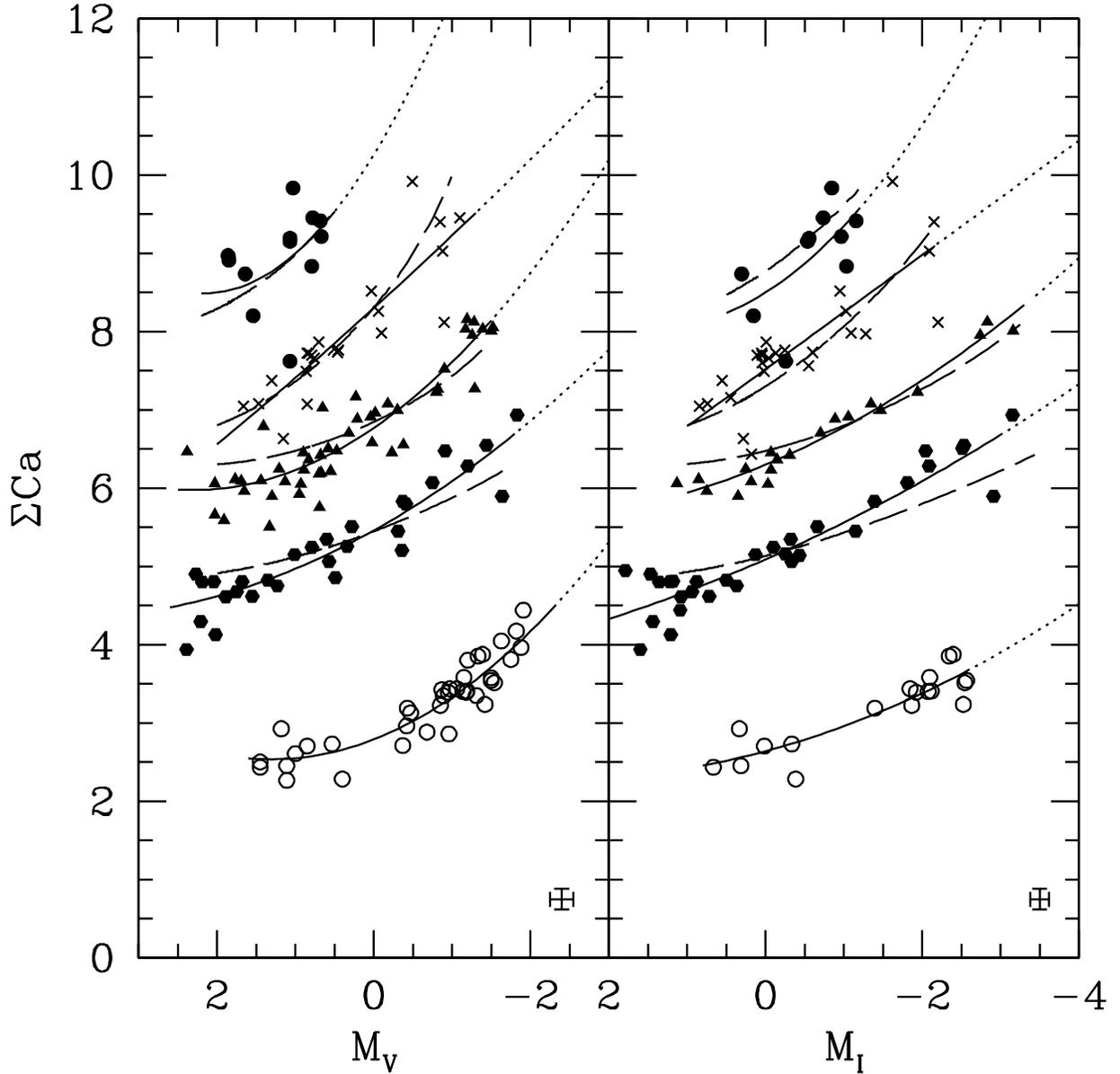}
\caption{Stars in the M$_V$--$\Sigma Ca$ and M$_I$--$\Sigma Ca$ planes for
the clusters in which we have observed stars along the RGB: NGC 7078
(open circles), NGC 288 (hexagons), NGC 104 (triangles), NGC 2141
(crosses) and NGC 6791 (filled circles). The individual quadratic fit
to each cluster is plotted (solid lines). Dotted lines represent the
extrapolation of the fit in the magnitude range where there are no
calibration stars. We also plotted the theoretical predictions for
each of them (dashed lines). The models have been shifted to match
approximately the cluster sequences (see text for details). Errorbars
are omitted for clarity, but the typical error is shown on the lower
rigth corner.\label{luminosityfig}}
\end{figure}

\placefigure{luminosityfig}

The next step is to relate the CaT index with metallicity. The
strength of the absorption lines mainly depends on the chemical
abundance, stellar effective temperature (T$_{eff}$) and surface
gravity (log $g$). Therefore, to relate the equivalent width of the
CaT lines with metallicity it is necessary to remove the T$_{eff}$ and
log $g$ dependence. \citet{adc91} and \citet{ol91} demonstrated that
the cluster stars define a sequence in the Luminosity--$\Sigma Ca$
plane, using luminosity measures from indicators like M$_I$ or
(V-V$_{HB}$). These sequences are separated as a function of the
cluster metallicity. The theoretical explanation of this can be found
in \citet{pont04}, using \citet{jcj92} models, which describe the
behaviour of the CaT lines as a function of T$_{eff}$, log $g$ and
metallicity.

It is necessary to study the morphology of the sequence defined by
each cluster in the Luminosity--$\Sigma Ca$ plane. From a theoretical
point of view, the increment of luminosity along the RGB comes with a
drop in T$_{eff}$ and log $g$ that decreases and increases the
strength of the lines, respectively. The result is a modest increment
in $\Sigma Ca$ with luminosity ($\delta\Sigma Ca/\delta
M_I\sim$0.5). Moreover, the models predict that $\Sigma Ca$ increases
more rapidly with luminosity in the upper part of the RGB (above the
HB) than in the lower part.  In other words, the sequence defined by
each cluster might not be linear and might be best described adding a
quadratic component. The \citet{jcj92} models also predict that
$\Sigma Ca$ increases more rapidly when log $g$ decreases, or when the
luminosity increases, for the more metal-rich clusters than for the
more metal-poor ones. Therefore, the linear and quadratic terms, which
characterize the sequence defined for each cluster in the
luminosity--$\Sigma Ca$ plane, increase with metallicity, as can be
seen in Figure 15 of \citet{pont04}.

Observationally, the variation in $\Sigma Ca$ with metallicity has
traditionally been studied from (V-V$_{HB}$), which removes any
dependence on distance and reddening
\citep[e.g.][]{adc91,r97a,c04}. In this context, it is found that
clusters define linear sequences in the (V-V$_{HB}$)--$\Sigma Ca$
plane, where the reduced equivalent width, $W'$, is defined as $\Sigma
Ca=W'_{HB}$+$\beta$(V-V$_{HB}$). \citet{r97a} found that the slopes
of these sequences were the same for all clusters in their sample,
independently of their metallicity. Therefore only $W'_{HB}$ changes
from one cluster to another, and its variation is directly related to
metallicity. Other studies have reached the same conclusion using open
and globular clusters \citep[e.g.][]{ol91}. \citet{pont04} \citep[see
also][]{adc91} have demonstrated that this also occurs in the
M$_V$-$\Sigma Ca$ and M$_I$-$\Sigma Ca$ planes. However, no studies
have observed the theoretical predictions that cluster sequences are
not exactly linear with luminosity, or that their shape depends on
metallicity.

The main objective of this study is to apply the relationships
obtained to derive metallicities of individual stars in Local Group
galaxies, which in general have had multiple star formation epochs and
do not always have a well defined HB (e.g. LMC: Carrera et al. 2007;
SMC: No\"el et al. 2007; Leo A: Cole et al. 2007). For example, the
Magellanic Clouds do not have a measurable HB in the CMD, and in
studies which define the reduced equivalent width as a function of
$(V-V_{HB})$ (\citep[e.g.][]{c05}), the HB position has been taken as
that of the red-clump. However, in the Magellanic Clouds, the position
of the red-clump is about 0.4 magnitudes brighter than the HB. This
only implies underestimating the metallicity by $\simeq$ 0.15 dex,
which is similar to the uncertainty on the metallicity determination
itself. Distances to Local Group galaxies are in general determined
with an accuracy greater than 0.4 mag., and so, even if the error on
the derived metallicity due to the uncertainty in the position of the
HB is not large, it can be minimized by defining the reduced equivalent
width as a function of absolute magnitude. This point is also
important in the case of open clusters, which hardly ever have a
HB or, if they do, it is usually not well defined.  For
this reason, like \citet{pont04}, we redefine $W'$ as the value of
$\Sigma Ca$ at M$_V$=0 (hereafter $W'_V$) or M$_I$=0 (hereafter
$W'_I$).

First we will study in detail the morphology of the cluster sequences
in the Luminosity--$\Sigma Ca$ plane. As discussed above, from a
theoretical point of view, we expect that these sequences are not
exactly linear. We have observed stars along the RGB in 5 clusters
covering the whole metallicity range. In Figure \ref{luminosityfig} we
have plotted stars observed in these clusters in the M$_V$--$\Sigma
Ca$ and M$_I$--$\Sigma Ca$ planes. These stars have magnitudes in the
ranges -2$\leq$M$_V\leq$2 and -3$\leq$M$_I\leq$2 (or
-2.3$\leq$V-V$_{HB}\leq$1.8). These ranges contain both stars brighter
and fainter than previous works \citep[e.g.][]{r97a,c04}. Note that
the strength of the CaT lines increases more rapidly in the upper part
of the RGB, as predicted by \citet{pont04} using \citet{jcj92}
models. These observations can be used to obtain a new relationship
between $\Sigma Ca$, absolute magnitude and metallicity valid for all
the stars in the RGB, that takes into account the curvature in the
Luminosity--$\Sigma Ca$ plane.  The sequence of each cluster has been
fitted with a quadratic function such that $\Sigma
Ca$=$W'_{V,RGB}$+$\beta$M$_V$+$\gamma$M$_V^2$. We plotted the result
when the stars of each cluster are fitted independently in
Figure~\ref{luminosityfig}. The coefficients of the fit are shown in
Table \ref{luminositytable}. From this, it seems that $\beta$ tends to
increase with metallicity, as predicted theoretically. In the case of
$\gamma$ this increment is not observed, i.e.  its variation does not
show a significant dependence on metallicity, except for the most
metal-rich cluster, which also has a large uncertainty.

\begin{figure}
\epsscale{1}
\plotone{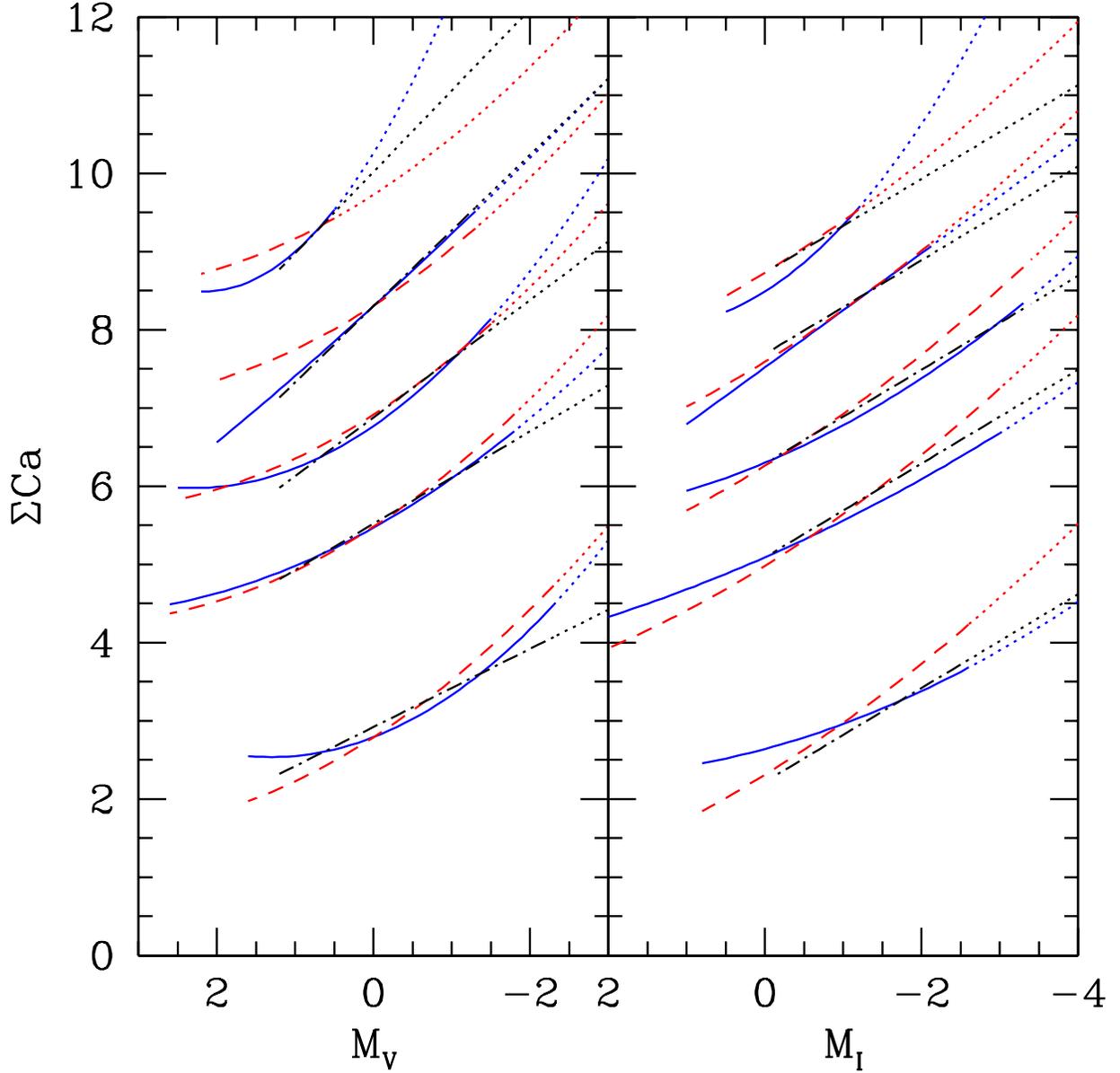}
\caption{Different fits to the sequences of the clusters in which we have 
observed stars along the RGB in the M$_V$--$\Sigma Ca$ and
M$_I$--$\Sigma Ca$ planes. Solid lines are the quadratic fit to each
cluster independently. Dashed lines are the quadratic fit when the
linear and quadratic terms are the same for all clusters. Finally,
dotted--dashed lines are the linear fits for stars brighter than
M$_V\leq$ 1.25 and M$_I\leq$ 0, assuming the same slope for all
clusters. Dotted lines are the regions in which the fits are
extrapolated. \label{luminositymodels}}
\end{figure}

Using the \citet{jcj92} empirical relations and the BaSTI stellar
evolution models \citep{pie04}, we have calculated theoretical
sequences for clusters with [Fe/H] $\geq -1$, which are plotted in
Figure \ref{luminositymodels} as dashed lines. These models were
obtained for [Fe/H] = +0.5, 0, $-0.5$ and $-1$, while the clusters
metallicities are [Fe/H] =+0.47, $-$0.14, $-$0.67 and $-$1.07
respectively. \citet{jcj92} did not compute relationships for more
metal-poor clusters. We used BaSTI isochrones with metallicities of
+0.32, $-0.28$, $-$0.58 and $-$0.98, respectively, in order to
estimate T$_{eff}$ and log $g$ along the RGB. The \citet{jcj92}
relationships were calculated for the two strongest CaT lines. To
compare the theoretical predictions with the observational sequences
we computed, using our own data, an empirical relation between $\Sigma
Ca_{8442+8662}$ obtained from these two lines and the $\Sigma Ca$ used
in this work, computed from the three CaT lines. We found is $\Sigma
Ca=0.13+1.21\Sigma Ca_{8442+8662}$. Applying this correction, we find
that the theoretical and observed cluster sequences still do not
match. There is a zero-point that changes from one cluster to another,
which is not surprising because the cluster metallicities are
not exactly the same as those used to compute the theoretical
relationships. Therefore, the theoretical sequences have been
shifted in order to superimpose them on the cluster ones. It can be
seen that models do not exactly reproduce the behaviour of the
observed cluster sequences. However, the prediction that the shape
changes from the metal-poor clusters to the metal-rich ones is
observed, although, as was mentioned before, these variations are
similar to the uncertainties.

We can simplify the problem if we assume that all clusters have the
same tendency, i.e. if we calculate a single slope and quadratic term
for the whole sample. So only the zero point changes among
clusters. To obtain these coefficients, we have performed an iterative
least-squares fit as described by \citet{r97a}. From a set of
reference values, we obtained the quadratic and linear terms of the
fit in iterative steps, until they converged to a single value within
the errors and allow only the zero point to change among clusters. The
values are: $\beta_V=-0.647\pm 0.005$ and $\gamma_V=0.085\pm0.006$. In
the same way, for M$_I$ we obtained $\beta_I=-0.618\pm 0.005$ and
$\gamma_I=0.046\pm 0.001$. In Figure \ref{luminositymodels} we have
plotted the individual fit for each cluster (solid line) and that when
the linear and quadratic terms do not change among clusters (dashed
lines). In both cases, the dotted lines represent the region where
there are no cluster stars and the fits have therefore been
extrapolated. As we can see in Figure \ref{luminositymodels}, in the
magnitude interval covered by cluster stars, both fits are similar and
give very similar values of $W'$ within the uncertainties. For
example, for NGC 7078, where the discrepancy is larger, we obtained
2.79 $\pm$ 0.06 and 2.79 $\pm$ 0.01 in $V$; and 2.64 $\pm$ 0.08 and
2.31 $\pm$ 0.01 in $I$, when the linear and quadratic terms change
among clusters or they are fixed, respectively. Larger differences
between both fits are found in the regions where the relationships are
extrapolated.

Moreover, in our case we are interested in measuring the strength of
the CaT lines in galaxies where we can observe only the upper part of
the RGB with a good S/N. The quadratic behaviour of the cluster
sequences in the Luminosity--$\Sigma Ca$ plane is not significant when
we observe stars with M$_I\leq$0 only (or M$_V\leq$1.25; this
magnitude limit has been selected in order to sample in both filters
the same number of stars in each cluster). For example, when we repeat
the previous procedure, but only for stars with M$_V\leq$ 1.25, we
find that the quadratic term is $\gamma_V$ = 0.004 $\pm$ 0.003, which
is negligible within the uncertainty. In the same way, when we only
observe stars with M$_V\geq$ 1.25 we obtain a similar result:
$\gamma_V=0.002\pm 0.01$. The same happens in the M$_I$--$\Sigma Ca$
plane, but here the quadratic terms are even smaller. According to
this, the cluster sequence can be considered linear above and below
M$_V$=1.25 and M$_I$=0, and we can fit it as $\Sigma$Ca =
$W'_V+\beta_V M_V$ or $\Sigma$Ca = $W'_I+\beta_I M_I$ on each side of
this point. Following the same iterative procedure as in the case of
the quadratic fit, we calculated the values of the slope $\beta$ for
M$_V\leq$ 1.25 and for M$_I\leq$ 0, obtaining $\beta_{V}=-0.74\pm0.01$
and $\beta_{I}=-0.60\pm0.01$, respectively. The linear fits for
M$_V\leq$1.25 and M$_I\leq$ 0 are represented in Figure
\ref{luminositymodels}, by dotted--dashed lines. In all cases, within
the ranges covered by the cluster stars, the linear fit to the bright
stars is equivalent, within the uncertainties, to the quadratic ones.

\begin{figure}
\epsscale{1}
\plottwo{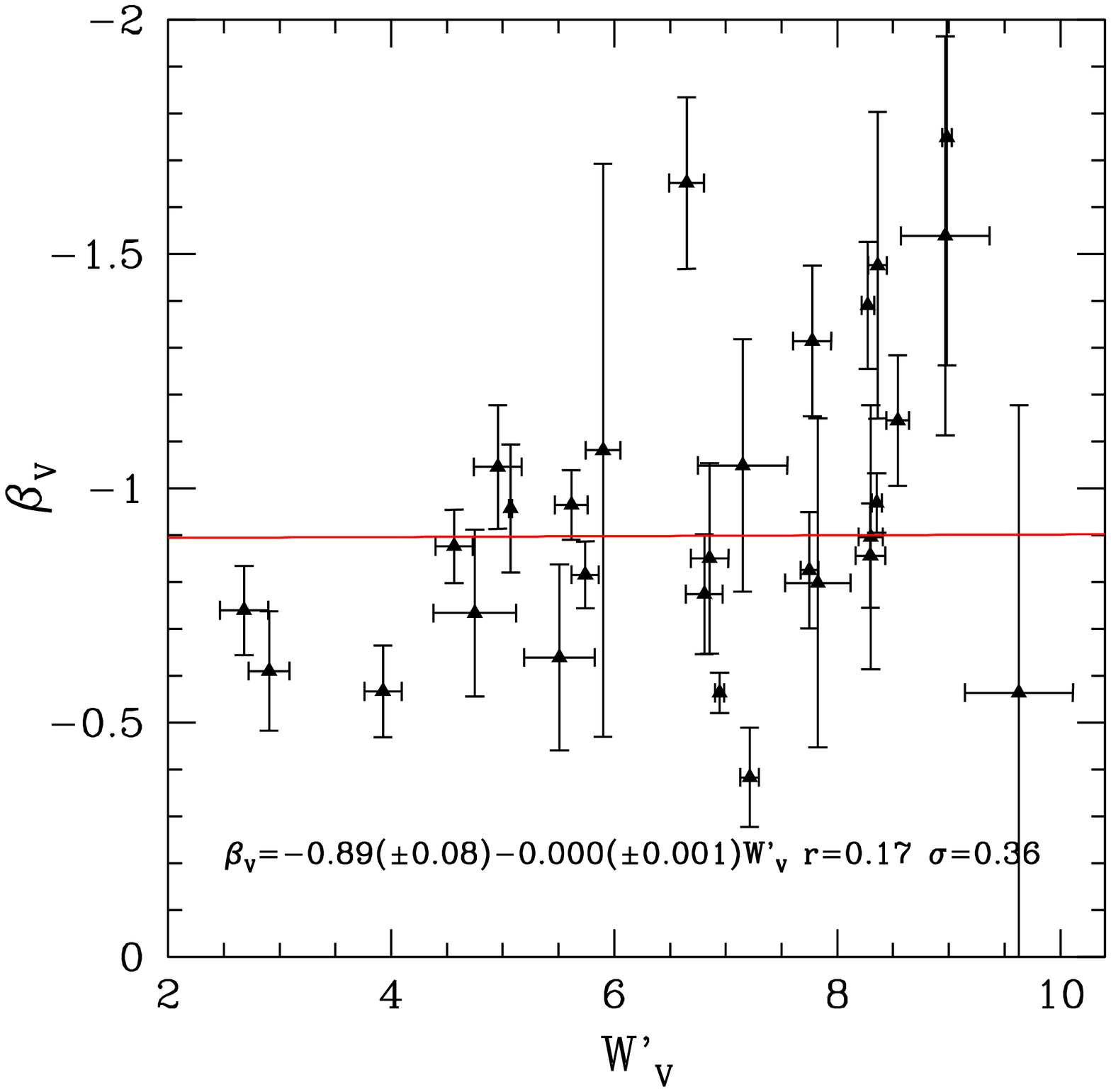}{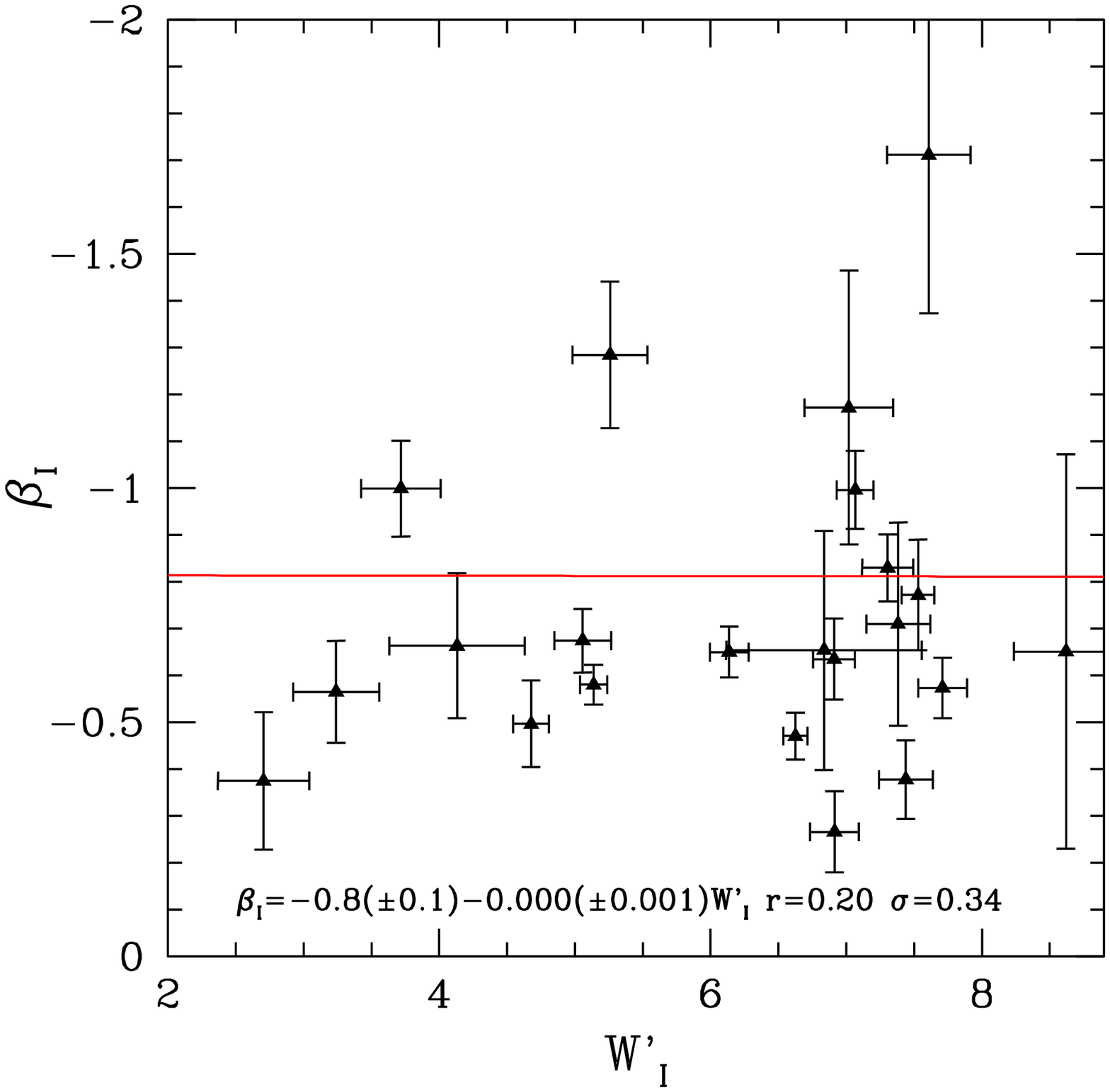}
\caption{Values of the slopes obtained from the individual fit for each 
cluster, versus $W'$. Solid lines are the linear fit, which is given
at the bottom. Note that there is no correlation between slope and
$W'$ (and therefore [Fe/H]) in any of the filters.\label{pendientefig}}
\end{figure}

Finally, for clusters where we have observed a wide range of
magnitudes we find that the slope ($\beta$) increases, although within
the uncertainties, with metallicity. We might check this point using
now all clusters in our sample. A total of 27 clusters in $I$ and 29
in $V$ have stars brighter than M$_I$=0 and M$_V$ = 1.25. We have
fitted the sequence to each cluster independently in the linear form
$\Sigma Ca=W'_{V,I}+\beta_{V,I} M_{V,I}$. The values obtained from the
slope have been plotted against $W'$, which is directly correlated with
metallicity, for each cluster in Figure~\ref{pendientefig}.  From this
figure it is seen that there is no significant relation between the
cluster slope and $W'$ (or [Fe/H]).  Therefore, from here on, we consider the
slope of the fit to be the same for the whole range of [Fe/H] and,
hence, for all objects.

In summary, as we are specially interested in obtaining metallicities
for stars in the upper part of the RGB with the CaT, where the
quadratic term is not significant and the slope can be fixed
independently of metallicity, we are going to use a linear fit with a
single slope for the calibration using the whole cluster sample. This
is what has been done in all previous calibrations of the CaT.

\begin{figure}
\epsscale{1}
\plotone{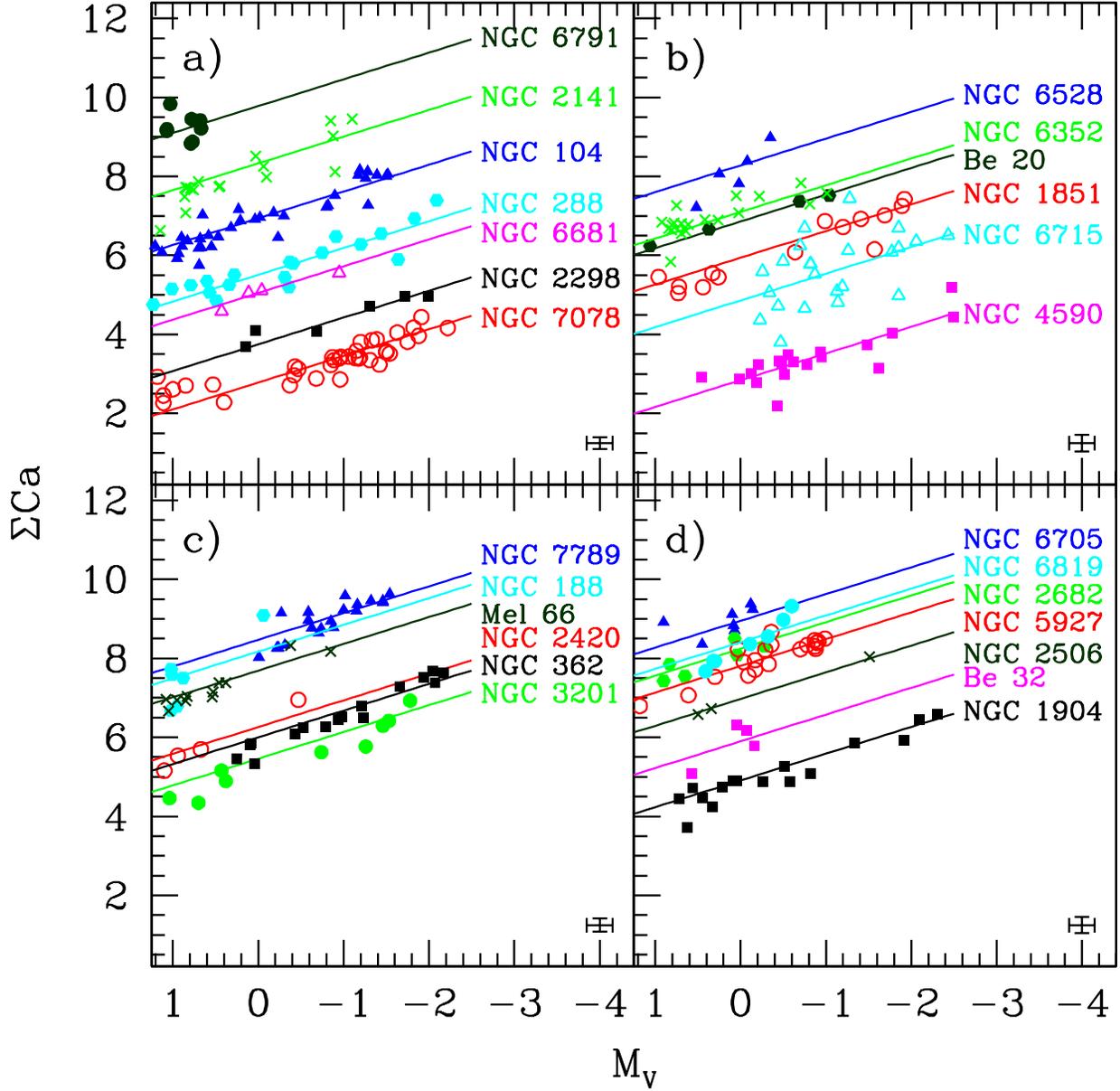}
\caption{Cluster sample in the M$_V$--$\Sigma Ca$ plane. Solid lines are 
the linear fit to the stars in each cluster when we assume that the
slope is the same for all of them. The typical error is shown on the
lower rigth corner.\label{mvsigmaca}}
\end{figure}

\begin{figure}
\epsscale{1}
\plotone{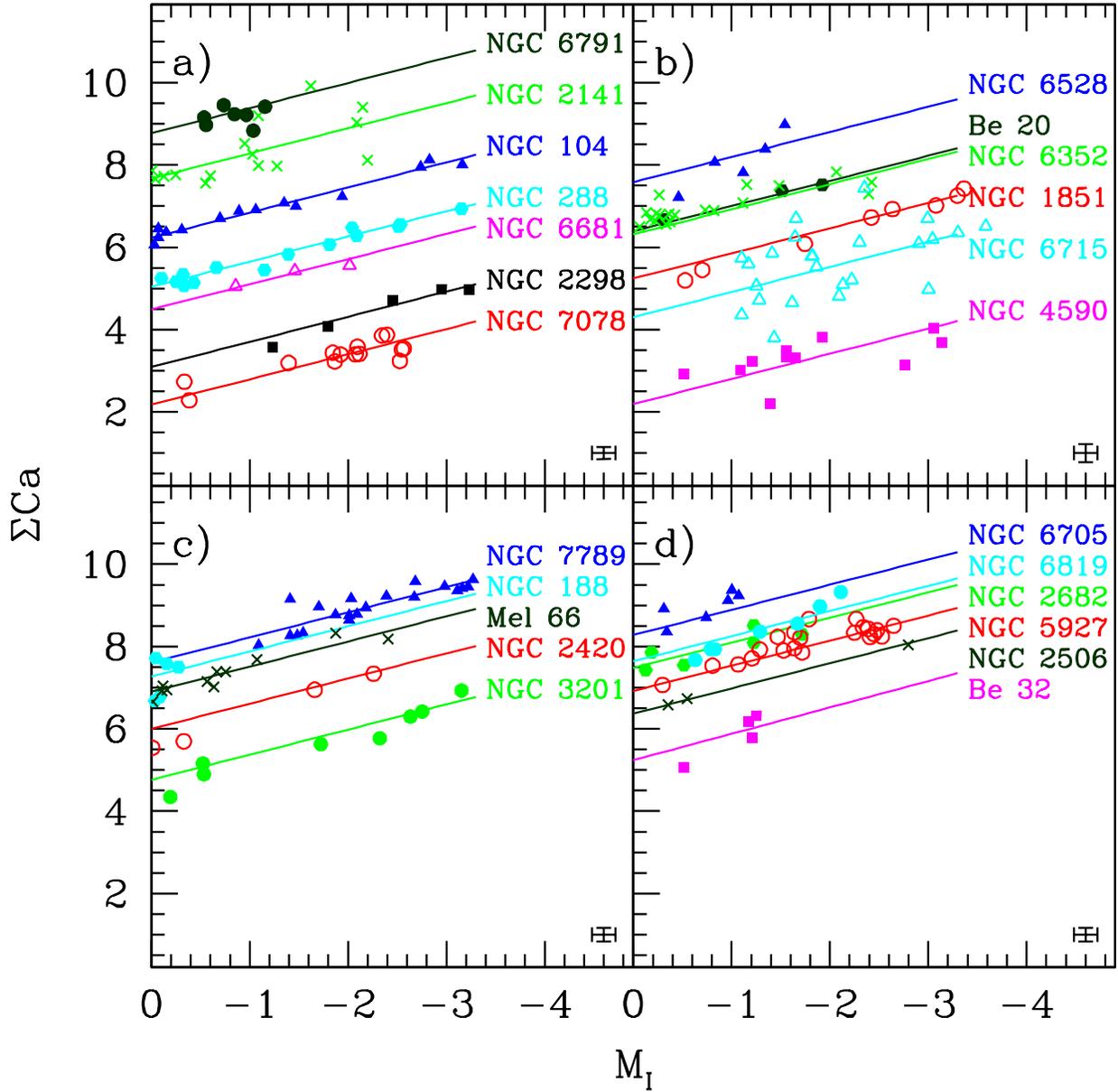}
\caption{Same as Figure \ref{mvsigmaca} but in the M$_I$--$\Sigma Ca$ plane.\label{misigmaca}}
\end{figure}

Figures \ref{mvsigmaca} and \ref{misigmaca} represent the clusters in
our sample in the M$_V$--$\Sigma Ca$ and M$_I$--$\Sigma Ca$ planes
respectively, together with the linear fit to each of them. Using the
same procedure as in the case of the quadratic fit discussed above, we
have obtained $\beta_V=-0.677\pm0.004$ and
$\beta_I=-0.611\pm0.002$\AA~mag$^{-1}$ in the M$_V$--$\Sigma Ca$ and
M$_I$--$\Sigma Ca$ planes, respectively. The value found in the
M$_I$--$\Sigma Ca$ plane is slightly larger than that obtained by
\citet{pont04}, $\beta_I=-0.48\pm0.02$ \AA~mag$^{-1}$. Although these 
authors used a different method to calculate the metallicity (they
fitted each cluster individually and obtained the mean of the slopes
of all of them), this is not the reason for the discrepancy because if
we follow the same procedure with our own data, again we find
$\beta_I=-0.61$.  There are no previous determinations of
$\beta_V$. The values obtained for $W'_V$ and $W'_I$ are listed in
Table \ref{fitstable}.
 
\placefigure{luminositymodels}

\placefigure{mvsigmaca}

\placefigure{misigmaca}

\section{The Ca {\footnotesize II} Triplet metallicity scale\label{catmetallicityscale}}

An important point in this study is the reference metallicities. It
would be ideal to use the same metallicity scale for both open and
globular clusters, and that this would have been obtained from
high-resolution spectroscopy. In the literature we can find two
globular cluster metallicity scales obtained from high resolution
spectroscopy: \citet[hereafter CG97]{cg97} and \citet[hereafter
KI03]{ki03}. There is a third metallicity scale obtained from
low-resolution data: \citet[hereafter ZW84]{zw84}. There are
systematic differences among these three scales, but there is no
reason to prefer any particular one of them. For this reason, here we
are going to study the behaviour of the CaT lines with metallicity in
these three scales. Lamentably, there is not a homogeneous metallicity
scale obtained from high-resolution spectroscopy for open
clusters. However, the metallicities of some of them have been
obtained directly in the CG97 scale by some authors: NGC 6819
\citep{bra01}; NGC 2506 \citep{cbgt04}; NGC 6791 \citep{gratton06} and 
Berkeley 32 \citep{sestito06}. These metallicities were obtained using
Fe {\footnotesize I} and Fe {\footnotesize II} lines. For the other 8
open clusters in our sample there are also metallicities obtained from
high-resolution spectroscopy in RGB stars and using Fe
{\footnotesize I} and Fe {\footnotesize II} lines in a similar way to
CG97. Even though some discrepancies could exist because the
procedures are not exactly the same, we are considering these
metallicities also to be on the CG97 scale. The reference values in
this scale are listed in column 2 of Table \ref{clustersample} and the
sources for each of them are listed in column 3. The reference
metallicities in the ZW84 and KI03 are listed in columns 4 and 5
respectively. In both cases, we have used only values obtained
directly by these authors.

\subsection{Calibration in the CG97 metallicity scale\label{cg97calibration}}

\begin{figure}
\epsscale{1}
\plotone{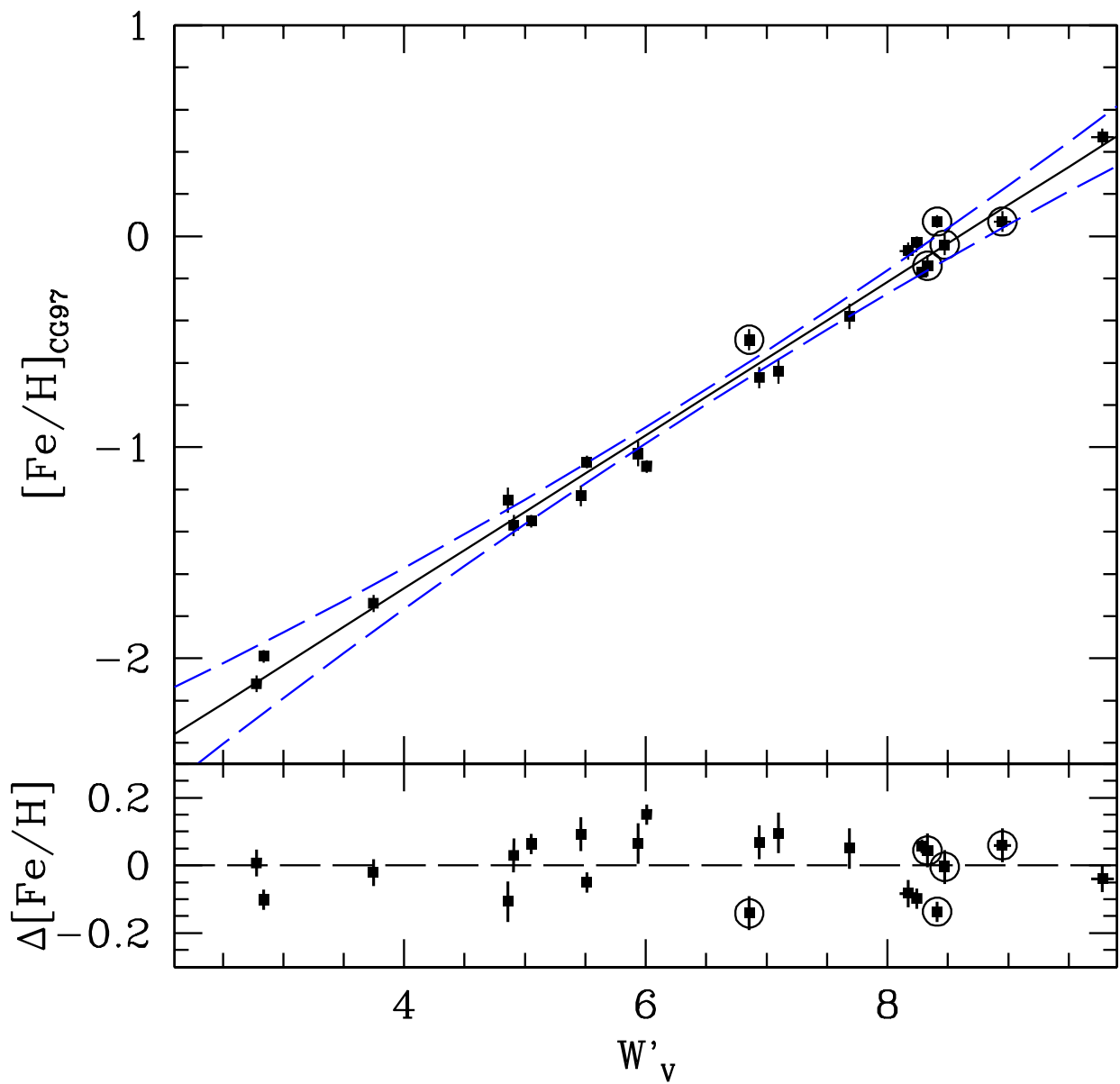}
\caption{Top panel: [Fe/H] versus $W'_V$. The solid lines are the best 
linear fit to the data. Dashed lines define the confidence band of the
fit. Open circles are clusters younger than 4 Gyr. The residuals of
the linear fit are shown in the bottom panel. Note that the $W'_V$
errors are smaller that the size of points in most cases. The clusters excluded
from the analysis (NGC 2420, NGC 2506 and Berkeley 32) have not been plotted.\label{calv}}
\end{figure}

\placefigure{calv}

\begin{figure}
\epsscale{1}
\plotone{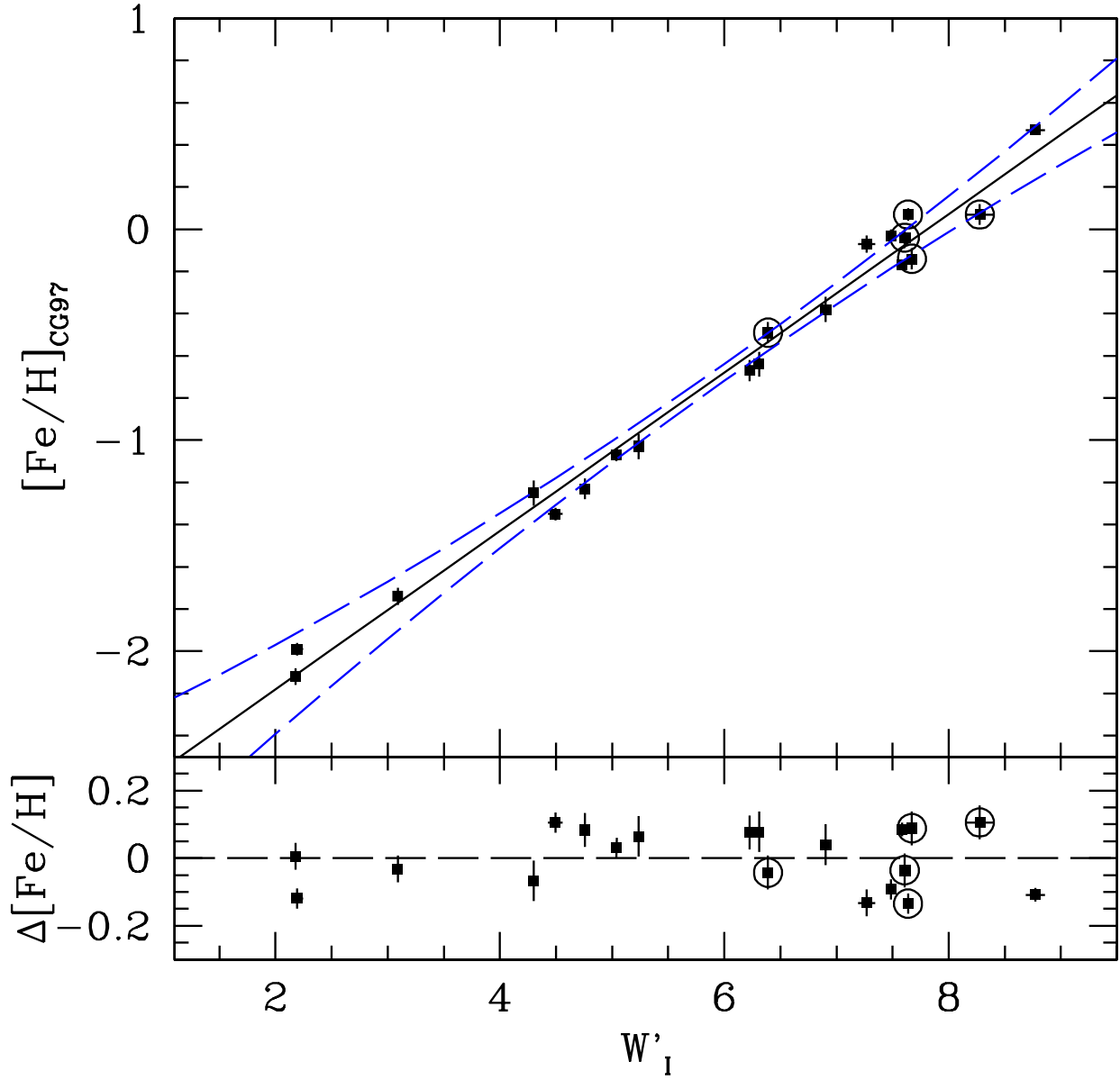}
\caption{Same as Figure \ref{calv} but with  $W'_I$. \label{calI}}
\end{figure}

Figures \ref{calv} and \ref{calI} show the run of $W'_V$ and $W'_I$
with metallicity. In most cases, the errors are smaller than the size
of the points. The circles indicate clusters younger than 4 Gyr. The
solid line shows the best fit to the data. The dashed lines represent
the 90\% confidence level. Note that in both cases there is a linear
correlation. The bottom panels show the residuals of the linear
fit. We have used 22 clusters for the calibration in $V$ and 20 for
that in $I$. There are three clusters that differ from the fit by more
than 0.2 dex in both filters. These clusters are NGC 2420, NGC 2506
and Berkeley 32. They have been excluded from the analysis. In the
case of NGC 2420, only 6 stars in $V$ and 4 in $I$ are radial velocity
members. This, together with a relatively large uncertainty in its
metallicity \citep{gratton00}, contributes to its large error bar. In
the case of NGC 2506 and Berkeley 32, there are only 3 and 4 stars
respectively with membership confirmed by their radial
velocities. Thus, slight differences in the $\Sigma Ca$ value of one
of them could change the derived $W'$ significantly. Two of the three very deviant clusters (NGC 2420 and NGC 2506) have ages
less than 4 Gyrs, but 5 other young clusters fit the mean relationships in
Figures \ref{calv} and \ref{calI} to better than 0.2 dex. We doubt therefore that cluster
age is the major cause of the large deviations. 

\placefigure{calI}

The best linear fits shown in Figures \ref{calv} and \ref{calI}, are:

\begin{equation}
[Fe/H]_{CG97}^V=-3.12(\pm0.06)+0.36(\pm0.01)W'_V~~\sigma_V=0.08\label{cg97v}
\end{equation} 

\begin{equation}
[Fe/H]_{CG97}^I=-2.95(\pm0.06)+0.38(\pm0.01)W'_I~~\sigma_I=0.09\label{cg97i}
\end{equation} 

Some studies have predicted that this relationship may present a
curvature due to the loss of CaT index sensitivity at high
metallicities \citep[e.g.][]{dtt89}. \citet{c04} investigated this
point adding a quadratic term. They found that the coefficient of this
term is insignificant and does not improve the quality of the fit. We
performed the same analysis in our sample, which covers a wider range
of ages and metallicities, finding a similarly insignificant influence
of a quadratic term.

\subsection{Calibration on Other Metallicity Scales\label{othermetallicities}}

\begin{figure}
\epsscale{1}
\plotone{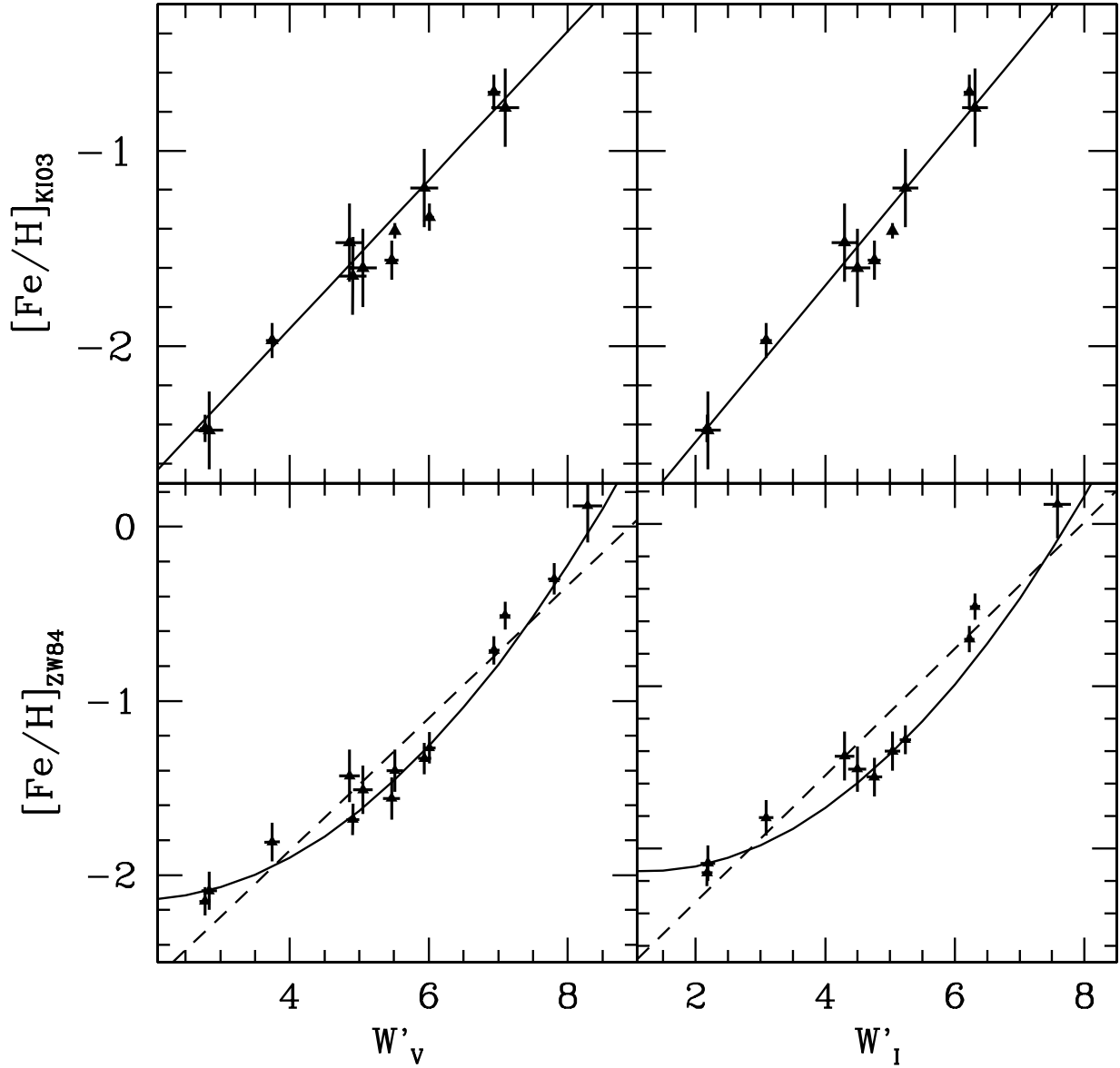}
\caption{$W'_V$ (left) and $W'_I$ (right) versus [Fe/H] on the KI03 (top) and the ZW84 (bottom) metallicity scales. The lines are the best
fit to the data. In the case of the ZW84 metallicity scale, a second-order polynomial results in an improvement
of the fit.\label{metallicities}}
\end{figure}

In this section we study the behaviour of the CaT on the ZW84 and KI03
scales. In Figure \ref{metallicities} we have plotted the
metallicities in ZW84 (bottom) and KI03 (top) listed in Table
\ref{clustersample} versus $W'_V$ (left) and $W'_I$ (right), respectively. 

In the case of the KI03 metallicity scale (top panels), the behaviour
of $W'$ with metallicity is linear, as for the CG97 scale. These
authors used three stellar atmosphere models to obtain
metallicities. For simplicity, in Figure \ref{metallicities} we have
plotted only the metallicity values obtained using MARCS
models. However, a linear behaviour is also found when we use the
metallicities computed from the Kurucz models with or without
convective overshooting. The linear fits for each of the three models
are:

\begin{mathletters}
\scriptsize
\begin{eqnarray}
&[Fe/H]_{KI03}^V&=-3.42(\pm0.03)+0.37(\pm0.01)W'_V~\sigma=0.10~(MARCS) \label{ki03vm}\\
&[Fe/H]_{KI03}^V&=-3.43(\pm0.03)+0.38(\pm0.01)W'_V~\sigma=0.10~(Kurucz~with~convective~overshooting) \\
&[Fe/H]_{KI03}^V&=-3.51(\pm0.03)+0.40(\pm0.01)W'_V~\sigma=0.10~(Kurucz~without~convective~overshooting) 
\end{eqnarray}
\end{mathletters}

\begin{mathletters}
\scriptsize
\begin{eqnarray}
&[Fe/H]_{KI03}^I&=-3.29(\pm0.03)+0.40(\pm0.01)W'_I~\sigma=0.09~(MARCS)\label{ki03im} \\
&[Fe/H]_{KI03}^I&=-3.24(\pm0.03)+0.40(\pm0.01)W'_I~\sigma=0.09~(Kurucz~with~convective~overshooting) \\
&[Fe/H]_{KI03}^I&=-3.31(\pm0.03)+0.41(\pm0.01)W'_I~\sigma=0.09~(Kurucz~without~convective~overshooting)
\end{eqnarray} 
\end{mathletters}

Differences between metallicities derived with the MARCS
model and the models of Kurucz with or without overshooting are negligible.

This linear behaviour is not surprising because, as KI03 demonstrated,
their metallicities are linearly correlated with the CG97 values,
which are, at the same time, linearly correlated with our
$W'$. However, the metallicities calculated by KI03 are systematically
lower than the CG97 ones. KI03 studied this point and concluded that
the difference could be explained because they used different
T$_{eff}$ and log g values, as well as different atmosphere
models. The combination of all these can easily introduce systematic
differences in the globular cluster abundance scales.

In the case of ZW84, we have found that the data
are best fitted by a second-degree polynomial (solid line):

\begin{mathletters}
\begin{eqnarray}
&[Fe/H]_{ZW84}^V&=-1.98(\pm0.07)-0.18(\pm0.02)W'_V+0.05(\pm0.01)W'^2_V~~\sigma_V=0.10\\
&[Fe/H]_{ZW84}^I&=-2.07(\pm0.07)-0.12(\pm0.03)W'_I+0.05(\pm0.01)W'^2_I~~\sigma_I=0.09\label{zw84i}
\end{eqnarray} 
\end{mathletters}
 
In Section 4.3, we discussed several previous definitions and
measurement procedures of the CaT lines, and noted the loss of
sensitivity to the CaT lines strength in some cases (e.g. Armandroff
and Da Costa 1991) which also found a non-linear relationship between
the CaT index and metallicity. We mentioned that this non-linearity
was probably the result of the combination of a non-accurate
measurement of the CaT on strong-line stars and the particular
metallicity scale in use. In order to assess the relative importance
each factor, we will now compare the effects on the derived abundances
of alternatively i) assuming a linear relationship between $W'$ and
metallicity on the ZW84 metallicity scale and ii) adopting a Gaussian
to fit the CaT lines, which provides a poorer fit. When a linear
relationship between $W'_I$ and [Fe/H]$_{ZW84}$ is assumed, the derived
metallicity of a strong-line star, $W'_I$=8.5, is underestimated in
0.3 dex. In the case of a weak-line star, $W'_I$=2, again the
metallicity is underestimated in 0.2 dex. Similar results are obtained
when lines are not properly fitted. For example, as we saw in Section
\ref{catindex}, \citet{adc91} fitted the line profile with a Gaussian, 
resulting in that their index saturated for strong-line stars. The
relation between the reduced equivalent width obtained from their
index and metallicities in the CG97 scale is a second-degree
polynomial. If we then assume a linear relationship between this index
and [Fe/H]$_{CG97}$ for a strong-line star, its metallicity would be
underestimated in 0.3 dex. Similar result is obtained for a weak-line
star. We conclude therefore, that the effects on the derived
metallicity due to a poor fit to the line or the non-linearity of
the metallicity scale are comparable.

\subsection{The role of Age in the $W'_V$ ($W'_I$) versus [Fe/H] relationship}

\citet{pont04} investigated the influence of age in the $W'_V$ ($W'_I$)
versus [Fe/H] relationship from a theoretical point of view. They used
the theoretical calculations of CaT equivalent widths for different
values of $\log g$, T$_{eff}$ and metallicity calculated by
\citet{jcj92} together with the Padova stellar evolution models
\citep{girardi02}. They concluded that the variation of $W'$ with age
for a fixed metallicity would be negligible for clusters older than 4
Gyr.  However, this was not the case for the younger clusters. This is
observed clearly in Figure 15 by \citet{pont04}. For a given
metallicity, the sequences in the M$_V$--$\Sigma Ca$ and M$_I$-$\Sigma
Ca$ planes are separated as a function of their ages for clusters
younger than $\sim$4 Gyr. According to this calculation, for the same
metallicity, $W'$ decreases with age. Thus, metallicities for clusters
younger than 4 Gyr, calculated from calibrations computed from old
stars, will be underestimated. This age dependence is more important
in the M$_V$--$\Sigma Ca$ plane than in the M$_I$--$\Sigma Ca$
one. This means that $W'_I$ would be less sensitive to age than
$W'_V$.

Using the \citet{jcj92} models and the BaSTI stellar evolution models
\citep{pie04}, we have estimated the expected $W'$ differences as a
function of age. From these calculations, for two clusters with the
same metallicity and age 10.5 and 0.6 Gyr respectively, the youngest
cluster $W'_V$ would be approximately 0.7 \AA\ lower than that of the
oldest one. This implies that the metallicity obtained for young
clusters using this calibration would be 0.25 dex more metal-poor than
the actual metallicity. In the case of $W'_I$, the difference would be
0.4 \AA, so the metallicity obtained for young clusters would be 0.15
dex more metal-poor than the actual one. As we can see in Figure 15 by
\citet{pont04}, the difference would be similar for different
metallicities.
  
\begin{figure}
\epsscale{1}
\plotone{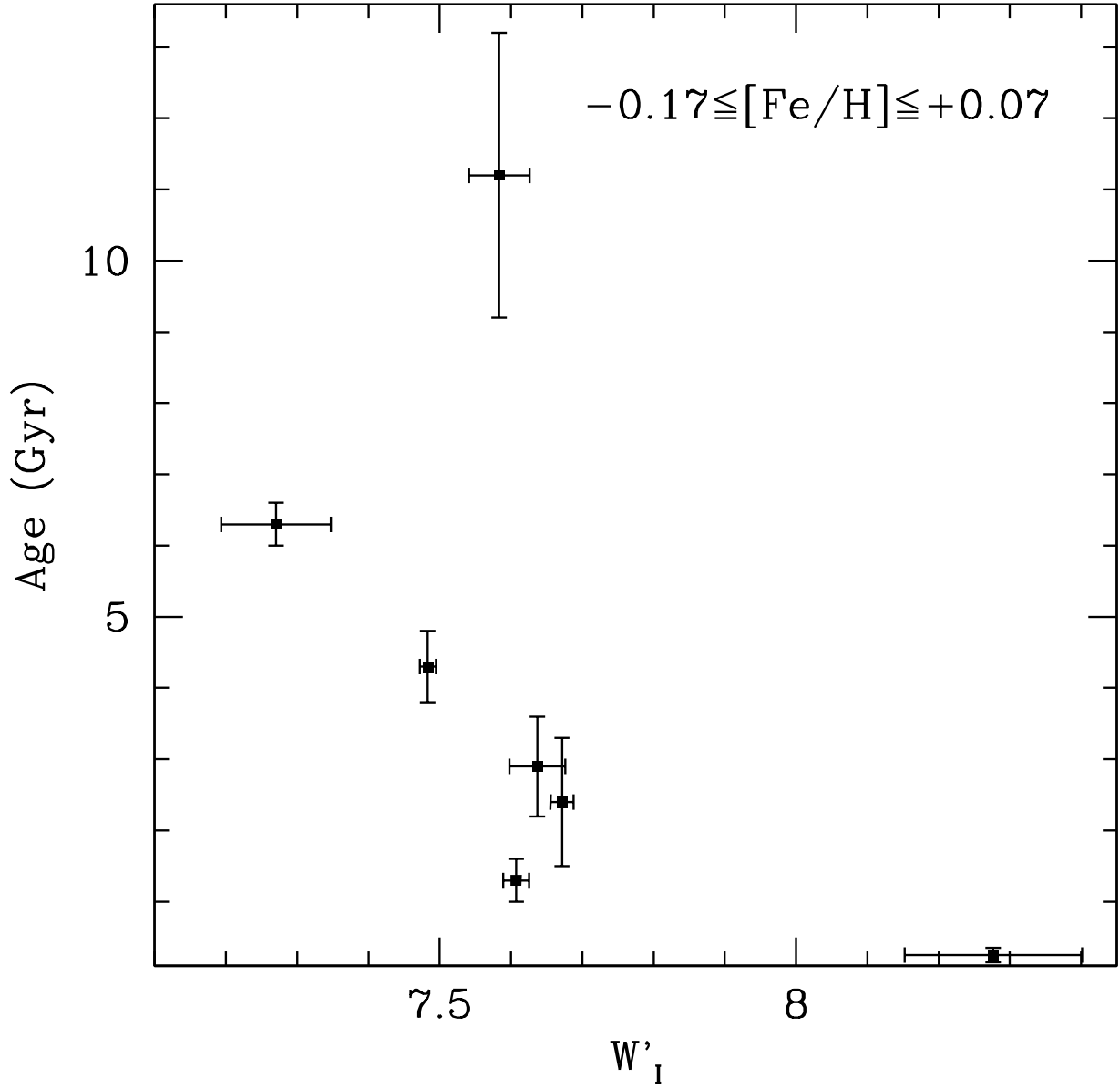}
\caption{$W'_I$ versus age for clusters with $ -0.17\leq$ [Fe/H] $\leq$ 
+0.07. Independently of their ages, all clusters have similar $W'_I$,
with the exception of the youngest cluster, NGC 6705 (0.25 Gyr). \label{agetest}}
\end{figure}

From our data, we confirm that the influence of age is weak. In Figure
\ref{agetest} we plot $W'_I$ versus age for clusters with $-0.17\leq$
[Fe/H]$_{CG97}\leq$ +0.07. We have selected this range because it
contains clusters with a wide range of ages and is small enough for
the metallicity differences to be within the uncertainties. We can see
that clusters with ages younger than 5 Gyr (NGC 2141, NGC 2682, NGC
6819 and NGC 7789) have similar $W'_I$ than the oldest one (NGC 6528).
There are only two clusters that deviate widely from the behaviour of
the others. One of these is the youngest cluster, NGC 6705, which has
a larger $W'_I$ than the oldest clusters. This is contrary to the
theoretical prediction that it should be smaller. However, we have to
take into account that differences of 0.5 \AA\ in $W'_I$ mean
differences of $\sim$0.1 dex in [Fe/H]. So the observed variations are
similar to the uncertainty in the determination of [Fe/H]. Our data
are not accurate enough to detect the influence of age because the
uncertainty in the metallicity determination of clusters is similar to
the expected variations due to age.

\placefigure{agetest}

\subsection{The influence of [Ca/Fe] abundance}

\begin{figure}
\epsscale{1}
\plotone{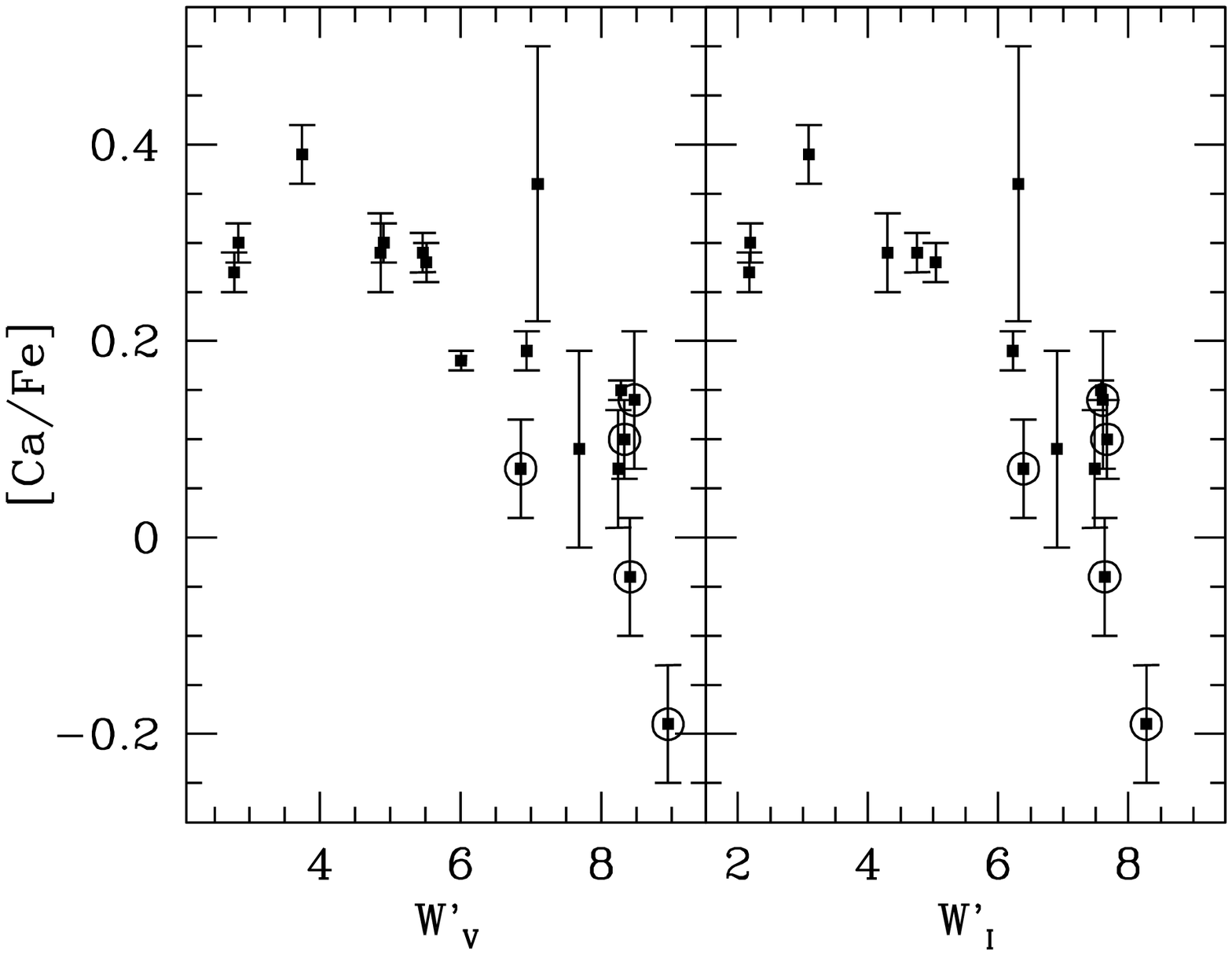}
\caption{[Ca/Fe] versus $W'_V$ (left) and $W'_I$ (right) for the clusters 
in our sample with Ca abundances available.\label{cafe}}
\end{figure}

The CaT has traditionally been used to infer Iron abundances from Ca
lines, and we also do so in this paper.  But, the CaT lines strength
should also be sensitive to the Ca abundances. In fact, the
relationships obtained in this work and those found in the literature
have been obtained assuming implicitly the specific relationship
between Ca and Fe followed by clusters used in the calibration (see
Figure \ref{cafe} for the relationship of the clusters used in this
work). Using these relationships to derive Fe abundances in stellar
systems with a different chemical evolution than the Milky Way,
reflected in the calibrating cluster sample, could give wrong
results.

In general, the relationship between the reduced equivalent width of
an atomic line and the chemical abundance of the corresponding element
is described by the curve of growth. This is only linear for very weak
and unsaturated lines. This is not the case for the CaT. As we can
find the [Ca/H] ratio for most of the clusters in our sample from the
literature, in Figure \ref{cafe} we have plotted $W'_V$ and $W'_I$
versus [Ca/H]. The relationship between both is equivalent to the
curve of growth. The relations obtained are:

\begin{eqnarray}
&[Ca/H]^V&=-2.51(\pm0.08)+0.30(\pm0.01)W'_V~~~\sigma=0.11\label{cahv}\\
&[Ca/H]^I&=-2.36(\pm0.08)+0.31(\pm0.01)W'_I~~~\sigma=0.11\label{cahi}
\end{eqnarray} 

As in the case of the [Fe/H] relationship, we obtain a linear
dependence. However, note that in this case the errors of the fit are
larger. This may be related to the inhomogeneity of the [Ca/H]
abundances, which were obtained from different sources.

In any case, even though [Ca/H] changes linearly with $W'$, [Fe/H]
does not have to do likewise. However, as we see in Figures \ref{calv}
and \ref{calI}, the relationship between [Fe/H] and $W'$ is also
linear. On the other hand, since the [Ca/H] and [Ca/Fe] abundances are
related according to [Fe/H] = [Ca/H] $-$ [Ca/Fe], we can expect that
[Ca/Fe] also changes linearly with $W'$ (and with [Fe/H]), if the
relation with [Ca/H] is linear. In fact, in Figure \ref{cafe} we can
check that this is the case over the whole range of [Fe/H] except for
the most metal-poor clusters. Note however that the linear behaviour
of $W'$ with [Ca/H] and [Ca/Fe] is a characteristic of our particular
sample, but this would not have to be the rule.

The problem of the relation between the CaT, [Ca/H] and [Fe/H] has
been addressed by \citet{idiart97} from an empirical point of
view. For their sample of late-type stars (G and K), they found that
the dominant stellar parameter controlling the behaviour of the CaT
lines is metallicity, and contrary to what would be expected, the
[Ca/Fe] ratio has practically no effect on the CaT index. However, all
the stars in their sample follow the same relationship between Ca and
Fe, so they cannot check in a general way the influence of the
[Ca/Fe] ratio.

To properly investigate the influence of the [Ca/Fe] ratio, it is
necessary to have objects with the same metallicities and different
[Ca/H] ratios. In our sample, most of the metal-poor clusters have
high $\alpha$-element abundances relative to Fe, as is the case for
Ca. On the other hand, open clusters are metal-rich and have low
$\alpha$-element abundances. To study the influence of the [Ca/Fe]
ratio on the CaT calibration as a function of metallicity it would be
necessary to include metal-rich objects with high $\alpha$-element
abundances (i.e.\ stars in the Milky Way bulge) and metal-poor objects
with low $\alpha$-element abundances (i.e.\ perhaps stars in dwarf
galaxies). This sort of work would need a huge observational effort,
which explains why it has not been done until now.

\placefigure{cafe}

\section{Derived cluster Metallicities\label{derivedmetallicities}}

We will use the relationships derived in previous sections to estimate
the metallicities in the three observed clusters without previous
determinations. In fact, we have observed Collinder 110, a poorly
studied cluster with no previous spectroscopic metallicity
determinations. For Berkeley 39, only \citet{f02} have determined its
metallicity from low-resolution spectroscopy. The sequences of these
clusters in the M$_I$--$\Sigma Ca$ plane have been plotted in Figure
\ref{msigmacatest}.

\placefigure{msigmacatest}
\subsection{Berkeley 39}

The first colour--magnitude diagram of this open cluster was published
by \citet{kaluzny89}. These authors calculated a distance modulus of
(m-M)$_V$ = 13.4 and $E(B-V)=0.12$. These values agree with the
determinations of \citet{carraro94}, who also used colour--magnitude
diagrams. The age of this cluster is 7 $\pm$ 1 Gyr \citep{swp04}.

There are few determinations of its metallicity. From photometric data
\citet{twarog97} estimated [Fe/H] = $-0.18\pm 0.03$, while from
low-resolution spectroscopy, \citet{f93} and \citet{f02} obtained
[Fe/H] = $-0.32 \pm 0.08$ and [Fe/H] = $-0.26\pm 0.09$
respectively. In our case we have 10 RGB stars which are cluster
members from their radial velocity, although only 5 stars have $I$
magnitudes available. Moreover, only 2 are brighter than M$_I$ = 0;
nevertheless, the other 3 have magnitudes close to this value. We
therefore used all 5 stars. From Equation \ref{cg97i} we obtain
[Fe/H]$_{CG97}$ = $-0.14 \pm 0.02$. We have used the relationship as a
function of M$_I$ because the RGB is more resolved in the $I$ filter,
and this relation is less sensitive to age.  The calculated value is
slightly more metal-rich than previous spectroscopic
determinations. In the KI03 and ZW84 scales we obtain [Fe/H]$_{KI03}$
= $-0.33 \pm 0.14$ and [Fe/H]$_{ZW84}$ = $-0.23 \pm 0.25$ respectively
from Equations \ref{ki03im} and
\ref{zw84i}. On these scales we have no young and/or metal-rich reference clusters, but, as we have checked before,
the influence of age is weak.

We have also calculated the radial velocity of this cluster. We find
V$_r$ = 59 $\pm$5 km s$^{-1}$, which is similar to values found
previously \citep[i.e.][V$_r$=55$\pm$7 Km s$^{-1}$]{f02}.

\subsection{Trumpler 5}

Trumpler 5, also named Collinder 105, is also a poorly studied
cluster, even though it was discovered about 75 yr ago. It is located
towards the Galactic anticentre in a rich star field in Monoceros, and
in a region of variable interstellar reddening. This has complicated
the studies of this cluster. In fact, only photometric studies could
be found in the literature \citep[e.g.][]{kaluzny98,kimsung03, pca04}
with the exception of the work by \citet{c04}, who observed the CaT
lines in a few stars on the RGB and derived the first spectroscopic
determination of its metallicity. The distance modulus and reddening
of this cluster have been derived from isochrone fitting. Most studies
converge on a reddening of $E(B-V)=0.6$
\citep[e.g.][]{kimsung03}. However, this does not happen in the case
of the distance, where the values lie between (m-M)$_0$ = 12.25
\citep{pca04} and 12.64 \citep{kimsung03}, corresponding to a distance
from the Sun of 2.4 or 3.4 kpc respectively. Also, the age and
metallicity have traditionally been estimated from isochrones. The age
of this cluster is estimated between 2.4 $\pm$ 0.2 \citep{kimsung03}
and 5.0 $\pm$ 05 Gyr \citep{pca04}, while the derived metallicity is
[Fe/H] = $-0.30 \pm 0.15$ dex \citep[e.g.][]{kimsung03,pca04}.

We have observed 21 stars in the field of Trumpler 5, 17 of which are
radial velocity members (Table~\ref{starsample}). The metallicity
derived from Equation \ref{cg97i} is [Fe/H]$_{CG97}$ = $-0.36 \pm
0.05$, which is more metal-rich (although within the error) than the
previous spectroscopic determination, [Fe/H] = $-0.56 \pm 0.11$, by
\citet{c04}. The alternative determination of the metallicity on the
KI03 and ZW84 scales gives [Fe/H]$_{KI03}$ = $-0.56
\pm 0.09$ and [Fe/H]$_{ZW84}$ = $-0.48 \pm 0.20$ respectively from
Equations~\ref{ki03im} and \ref{zw84i}

From our data we have also calculated the radial velocity of this
cluster, V$_r$ = 44 $\pm$ 10 km s$^{-1}$, which is similar to the
value derived by \citet[][V$_r$=54$\pm$5 Km s$^{-1}$]{c04}.

\subsection{Collinder 110}

Collinder 110 is a poorly populated cluster, even less studied than
Trumpler 5. Only two photometric studies can be found in the
literature for the last three decades. Using synthetic
colour--magnitude diagrams, \citet{bragagliatosi03} have estimated a
reddening of 0.38 $\leq E(B-V) \leq 0.45$ and distance modulus
(m-M)$_0$ between 11.8 and 11.9. From these values they derived an age
between 1.1 and 1.5 Gyr. Similar values were found by
\citet{dawson98}. There are no metallicity determinations for this
cluster in the literature. \citet{bragagliatosi03} tried to derive the
metallicity of this cluster from different stellar evolution models,
but concluded that the final result vary widely depending on the
models.

The metallicity derived from Equation \ref{cg97i} is [Fe/H]$_{CG97}$ =
$-0.01 \pm 0.07$. If we use Equations \ref{ki03im} and
\ref{zw84i} on KI03 and ZW84 metallicity scales we find [Fe/H]$_{KI03}$ 
= $-0.19 \pm 0.21$ and [Fe/H]$_{ZW84}$ = 0.00 $\pm$ 0.30. From our
data we can also provide the first determination of its radial
velocity, V$_r$ = 45 $\pm$ 8 km sec$^{-1}$.

\section{Summary\label{conclusions}}

We have observed the CaT lines in RGB stars in a sample of 29 clusters
of the Milky Way. This sample covers an age range of (13 $\leq$
Age/Gyr $\leq$ 0.25) and metallicity range of ($-2.2\leq$ [Fe/H]
$\leq$ +0.47). These are the widest ranges of ages and metallicities
in which the behaviour of the CaT has been investigated in a
homogeneous way until now. We have obtained relationships between the
CaT equivalent widths and metallicities on the scales of \citet{zw84},
\citet{cg97} and \citet{ki03}. The influence of other parameters, such
as age and [Ca/Fe] ratio, has been investigated. Moreover, for the
first time, the behaviour of the CaT lines as a function of luminosity
along the RGB has been studied for the whole range of metallicities in
our sample.

The main results of this work are:
\begin{itemize}

\item Theoretically, it has been predicted that the sequences of clusters in 
the Luminosity--$\Sigma Ca$ plane may not be linear, and that the
slope should change with metallicity. In this article we have
demonstrated that the nonlinear tendency and the change of the slope
can be (marginally) detected if a wide range of magnitudes in the RGB
is observed.

\item However, this behaviour is not significant if only the usual range 
of 3-4 magnitudes below the tip of the RGB is observed. For this
reason, for stars with M$_V\leq$ 1.25 or M$_I\leq$ 0, we have
considered that the sequences of the clusters in the M$_V$--$\Sigma
Ca$ and M$_I$--$\Sigma Ca$ planes are linear, and share a common slope,
independently of metallicity.

\item We have obtained relationships between the reduced equivalent 
width ($W'_V$ and $W'_I$) and metallicity on the \citet{zw84},
\citet{cg97} and \citet{ki03} scales. While on the \citet{cg97} and
\citet{ki03} scales these relationships are linear, in the case of the
\citet{zw84} scale, it is quadratic.

\item Theory predicts that the relationship between the CaT line equivalent 
widths and metallicity might be dependent on age, mainly for clusters
younger than 4 Gyr. We have studied the influence of age and found
that the expected differences due to age are similar to the
metallicity resolution of our work.

\item We have also investigated the influence of Ca abundances on the 
relationships between $W'_V$ and $W'_I$ and metallicity. We have found
that [Ca/H] also changes linearly with $W'_V$ and $W'_I$.

\item Finally, the  relationships obtained have been used to compute the 
metallicity of 3 clusters in our sample: Berkeley 39, Trumpler 5 and
Collinder 110. For the last one, there are no previous determinations
of its metallicity in the literature.

\end{itemize}

\acknowledgments

We warmly thank Dr. Antonio Aparicio for many fruitful discussions on
this paper, and a careful and critical reading of the
manuscript. Extensive use was made of the WEBDA database, maintained
at the university of Geneva, Switzerland. C.G. and R.C. acknowledge
the support from the Spanish Ministry of Science and Technology (Plan
Nacional de Investigaci\'on Cient\'{\i}fica, Desarrollo, e
Investigaci\'on Tecnol\'ogica, AYA2004-06343). E. P. acknowledge
support from the Italian MIUR (Ministero dell'Universit\`a e della
Ricerca) under PRIN 2003029437 entitled "Continuities and
discontinuites in the formation of the galaxy". R.Z. acknowledges the
support of the NSF under grant AST05-07364.  

Facilities:
\facility{VLT(FORS2)}, \facility{CAHA2.2m(CAFOS)},
\facility{CTIO4m(HYDRA)}, \facility{WHT(WYFFOS)},
\facility{WHT(ISIS)},
\facility{INT(IDS)}.




\clearpage


\clearpage

\clearpage

\clearpage

\clearpage

\clearpage

\clearpage

\clearpage

\clearpage

\clearpage

\clearpage

\clearpage

\clearpage

\clearpage

\clearpage

\clearpage

\begin{figure}
\epsscale{1}
\plotone{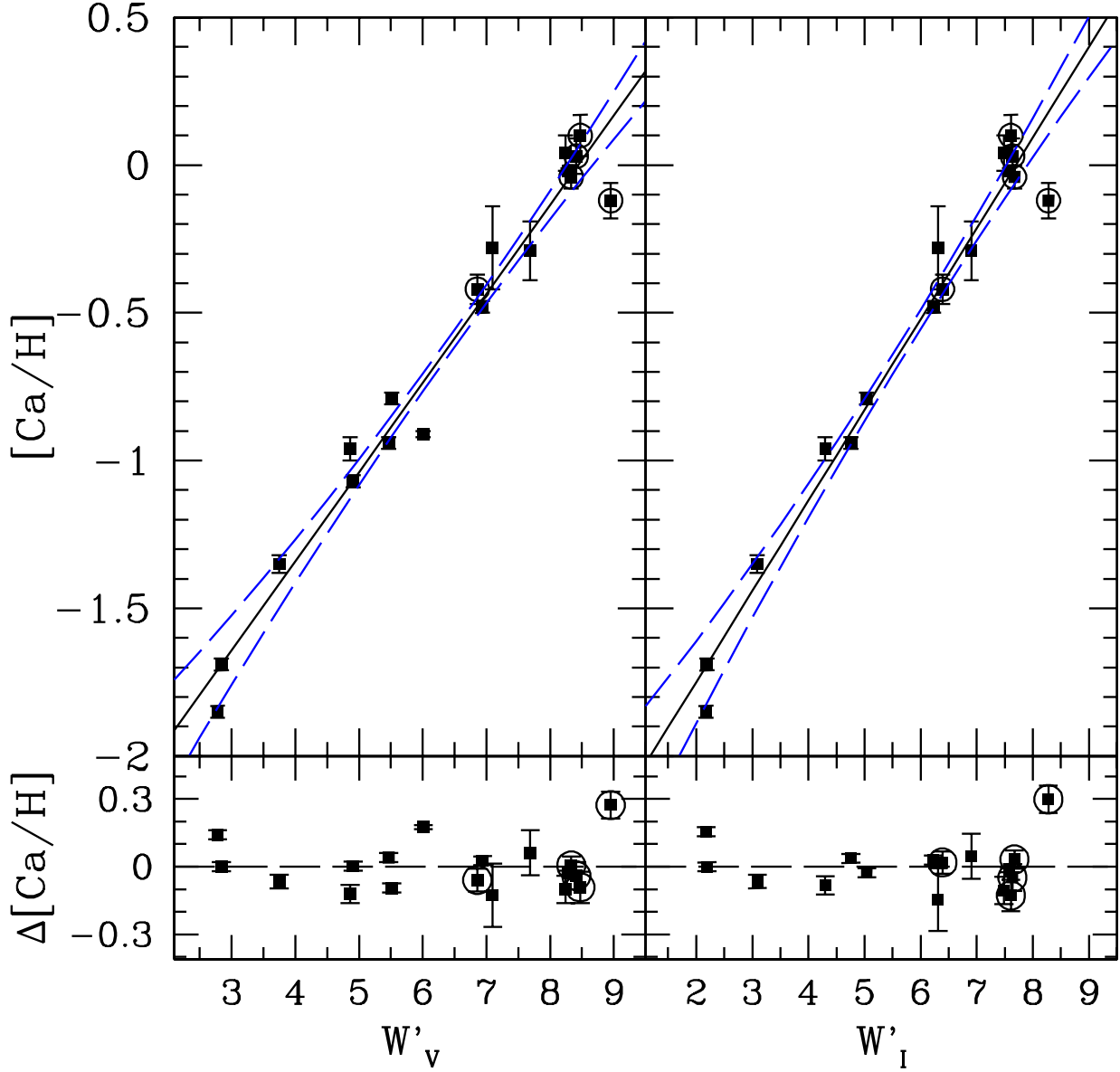}
\caption{$W'_V$ (left) and $W'_I$ (right) versus [Ca/H] ratio. The solid 
line is the best linear fit to the data.  As before, open circles are
clusters younger than 4 Gyr. The residuals of the linear fit are shown
in the bottom panel.\label{cah}}
\end{figure}
\clearpage

\begin{figure}
\epsscale{1}
\plotone{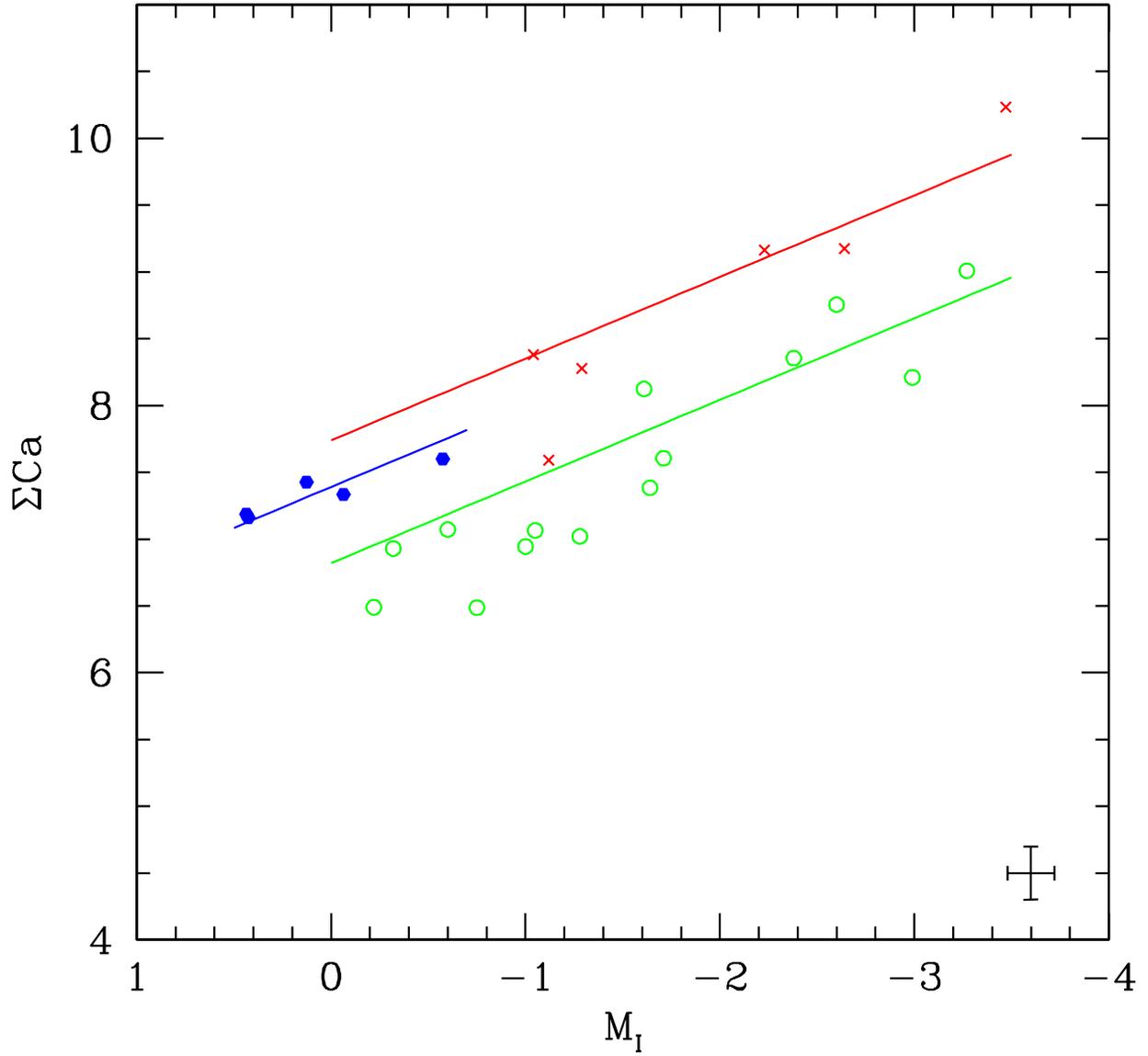}
\caption{Sequences of clusters Berkeley 39 (hexagons), Trumpler 5 (open 
circles) and Collinder 110 (filled squares) in the M$_I$--$\Sigma Ca$
plane. Solid lines are the linear sequence fits to the data for each
cluster when the same slope is assumed for all clusters. The typical
error is shown on the lower right corner.\label{msigmacatest}}
\end{figure}

\clearpage

\begin{deluxetable}{lcccccccccccc}
\tabletypesize{\scriptsize}
\rotate
\tablewidth{0pt}
\tablecaption{Cluster sample.
\label{clustersample}}
\tablehead{
\colhead{Cluster} & \colhead{[Fe/H]$_{CG97}$} & \colhead{Ref.}& \colhead{[Fe/H]$_{ZW84}$}&
\colhead{[Fe/H]$_{KI03}$} & \colhead{[Ca/H]} & \colhead{Ref.} &
\colhead{Age(Gyr)} & \colhead{Ref.} &\colhead{$(m-M)_V$} &
\colhead{$E(B-V)$} & \colhead{Ref.} & \colhead{Run}}
\startdata{shetrone00}
NGC 104 (47 Tuc) & $-0.67\pm0.03$ & 31 & $-0.71$ & $-0.70$ & $0.47\pm0.02$ & 31 & $10.7\pm1.0$ & 8 & 13.32 & 0.05 & 13 & 7 \\
NGC 188 & $-0.07\pm0.04$ & 2 & \nodata & \nodata & \nodata & \nodata & $6.30\pm0.3$ & 9 & 11.44 & 0.09 & 17 & 3 \\
NGC 288 & $-1.07\pm0.03$ & 1 & $-1.40$ & $-1.41$ & $-0.79\pm0.02$ & 33 & $11.3\pm1.1$ & 8 & 14.64 & 0.03 & 13 & 7 \\
NGC 362 & $-1.09\pm0.03$ & 1 & $-1.27$ & $-1.34$ & $-0.91\pm0.01$ & 33 & $8.7\pm1.5$ & 8 & 14.75 & 0.05 & 13 & 7 \\
NGC 1851 & \nodata & \nodata & $-1.36$ & \nodata & \nodata & \nodata & $9.2\pm1.5$ & 8 & 15.49 & 0.02 & 13 & 7 \\
NGC 1904 (M79) & $-1.37\pm0.05$ & 1 & $-1.69$ & $-1.64$ & $-1.07\pm0.02$ & 32 & $11.7\pm1.3$ & 8 & 15.53 & 0.01 & 13 & 7 \\
Berkeley 20 & $-0.49\pm0.05$ & 22 & \nodata & \nodata & $-0.42\pm0.05$ & 22 & $4.05\pm0.7$ & 9 & 15.84 & 0.38 &  22 & 7 \\
NGC 2141 & $-0.14\pm0.05$ & 22 & \nodata & \nodata & $-0.04\pm0.04$ & 22 & $2.45\pm0.9$ & 9 & 14.15 & 0.40 & 14 & 3,7 \\
Collinder 110 & \nodata & \nodata & \nodata & \nodata  & \nodata & \nodata & $1.3\pm0.2$ & 24 & 13.04 & 0.40 & 24 & 3 \\
Trumpler 5 & \nodata & \nodata & \nodata & \nodata & \nodata & \nodata & $5.7\pm2.3$ & 9 & 14.50 & 0.60 & 23 & 3,7 \\
NGC 2298 & $-1.74\pm0.04$ & 1 & $-1.85$ & $-1.64$ & $-1.35\pm0.03$ & 35 & $12.6\pm1.4$ & 8 & 15.54 & 0.13 & 13 & 7 \\
Berkeley 32 & $-0.29\pm0.04$ & 32 & \nodata & \nodata & \nodata & \nodata & $5.9\pm1.6$ & 9 & 12.85 & 0.08 & 25 & 3 \\
Melote 66 & $-0.38\pm0.06$ & 3 & \nodata & \nodata & $-0.29\pm0.10$ & 40 & $5.3\pm1.4$ & 9 & 13.63 & 0.14 & 26 & 7 \\
Berkeley 39 & \nodata & \nodata & \nodata & \nodata & \nodata & \nodata & $7.0\pm1.0$ & 9 & 13.24 & 0.11 & 26 & 7 \\
NGC 2420 & $-0.44\pm0.15$ & 3 & \nodata & \nodata & \nodata & \nodata & $2.2\pm0.3$ & 9 & 12.0 & 0.05 & 15 & 3 \\
NGC 2506 & $-0.20\pm0.01$ & 31 & \nodata & \nodata & \nodata & \nodata & $2.1\pm0.3$ & 9 & 12.60 & 0.09 & 16 & 3,7 \\
NGC 2682 (M 67) & $-0.03\pm0.03$ & 4 & \nodata & \nodata & $+0.01\pm0.06$ & 4 & $4.3\pm0.5$ & 9 & 9.65 & 0.04 & 17 & 3,7 \\
NGC 3201 & $-1.24\pm0.12$ & 1 & $-1.61$ & $-1.48$ & $-1.11\pm0.02$ & 34 & $11.3\pm1.1$ & 8 & 14.17 & 0.21 & 13 & 4 \\
NGC 4590 (M 68) & $-2.00\pm0.03$ & 1 & $-2.09$ & $-2.43$ & $-1.68\pm0.02$ & 34 & $11.2\pm0.9$ & 8 & 15.14 & 0.04 & 13 & 3,4,6,7 \\
NGC 5927 & \nodata & \nodata & $-0.30$ & \nodata & \nodata & \nodata & $10.9\pm2.2$ & 30 & 15.81 & 0.47 & 13 & 6 \\
NGC 6352 & $-0.64\pm0.02$ & 1 & $-0.51$ & \nodata & $-0.63\pm0.14$ & 36 & $9.9\pm1.4$ & 8 & 14.39 & 0.21 & 13 & 6 \\
NGC 6528 & $-0.17\pm0.02$ & 28,29 & $+0.12$ & \nodata & $-0.03\pm0.01$ & 37 & $11.2\pm2.0$ & 10 & 16.16 & 0.55 & 18 & 7 \\
NGC 6681 (M 70) & $-1.35\pm0.03$ & 1 & $-1.51$ & \nodata & \nodata & \nodata & $11.5\pm1.4$ & 8 & 14.93 & 0.07 & 13 & 7 \\
NGC 6705 (M 11) & $+0.07\pm0.05$ & 5 & \nodata & \nodata & $-0.12\pm0.06$ & 5 & $0.25\pm0.1$ & 11 & 12.88 & 0.43 & 11 & 2,5 \\
NGC 6715 (M 54) & \nodata & \nodata & $-1.42$ & $-1.47$ & $-0.98\pm0.04$ & 39 & $12\pm1.5$ & 27 & 17.77 &  0.16 & 19 & 7 \\
NGC 6791 & $+0.47\pm0.04$ & 6 & \nodata & \nodata & \nodata & \nodata & $12.0\pm1.0$ & 12 & 13.07 & 0.09 & 12 & 5 \\
NGC 6819 & $+0.07\pm0.03$ & 7 & \nodata & \nodata & $+0.03\pm0.06$ & 7 & $2.9\pm0.7$ & 9 & 12.35 & 0.14 & 20,7 & 1,5 \\
NGC 7078 (M 15) & $-2.12\pm0.04$ & 1 & $-2.15$ & $-2.42$ & $-1.88\pm0.02$ & 38 & $11.7\pm0.8$ & 8 & 15.31 & 0.09 & 13 & 7 \\
NGC 7789 & $-0.04\pm0.05$ & 4 & \nodata & \nodata & $+0.10\pm0.08$ & 4 & $1.3\pm0.3$ & 9 & 12.20 & 0.28 & 4,21 & 1,3 \\
\enddata
\tablerefs{
(1) \citet{cg97}; (2) \citet{hubbs90}; (3) \citet{gratton00}
(4) \citet{tautvaisiene05}; (5) \citet{gonzalezwallerstein00}; (6) \citet{gratton06};
(7) \citet{bra01}; (8) \citet{sw02}; (9) \citet{swp04};
(10) \citet{feltzingjohnson02}; (11) \citet{sung99}; (12) \citet{stetsonbg03};
(13) \citet{rosenberg99}; (14) \citet{carraro01}; (15) \citet{lee99};
(16) \citet{marconi97}; (17) \citet{sarajedini99}; (18) \citet{ortolani92};
(19) \citet{rosenberg04}; (20) \citet{rosvickvanderverg98}; (21) \citet{gim98};
(22) \citet{young05}; (23) \citet{kimsung03}; (24) \citet{bragagliatosi03}; 
(25) \citet{richtlersagar01}; (26) \citet{kassis97}; (27) \citet{laydensarajedini97}
(28) \citet{zoccali04}; (29) \citet{origlia05}; (30) \citet{fullton96}; (31) \citet{cbgt04}; (32)
\citet{sestito06}; (33) \citet{shetrone00}; (34) \citet{gratton89}; (35) \citet{mcwilliam92}; (36) \citet{gratton87}; (37) \citet{ccgb01};
(38) \citet{sneden97}; (39) \citet{brown99}; (40) \citet{gratton94}.}

\end{deluxetable}
\clearpage




\begin{deluxetable}{cccccc}
\tablecaption{Observing runs
\label{runs}}
\tablewidth{0pt}
\tablehead{
\colhead{Run} & \colhead{Date} & \colhead{Telescope} & \colhead{Instrument} & \colhead{Resolution}&
\colhead{\AA/pix}}
\startdata
1 & May 2002 & WHT & ISIS & 7000 & 0.41 \\
2 & April 2002 & WHT & WYFFOS & 4000 & 1.5 \\
3 & December 2002 & INT & IDS & 6000 & 0.45 \\
4 & January 2005 & CTIO 4m & HYDRA & 6000 & 0.9 \\
5 & June 2005 & CAHA 2.2m & CAFOS & 2000 &2.0 \\
6 & 2005 & VLT & FORS2 MXU & 5000 & 0.85 \\
7 & ESO Archive & VLT & FORS2 MXU/MOS & 5000 & 0.85 \\
\enddata
\end{deluxetable}

\clearpage
\begin{deluxetable}{lcccccccc}
\tablewidth{0pt}
\tablecaption{Star sample.
\label{starsample}}
\tablehead{
\colhead{Cluster} & \colhead{Id} & \colhead{$\Sigma$Ca} &
\colhead{$\sigma_{\Sigma Ca}$} & \colhead{V} &\colhead{I} &
\colhead{V$_r$} & \colhead{$\sigma_{V_r}$} & \colhead{Comments}}
\startdata
ngc104 & S2701 & 6.18 & 0.13 & 14.07 & 99.99 &  2.22 & 4.90 &  Member?\\
ngc104 & S2703 & 6.55 & 0.07 & 12.99 & 99.99 &  4.44 & 4.34 &  No member\\
ngc104 & S2705 & 7.27 & 0.02 & 12.08 & 99.99 &-13.24 & 3.68 &  \\
ngc104 & S2707 & 6.96 & 0.07 & 13.35 & 99.99 & -2.71 & 5.46 & \\
ngc104 & S2712 & 7.34 & 0.05 & 12.89 & 99.99 &  9.46 & 5.25 &  No member\\
\enddata
\tablerefs{{\bf NGC 104} \citet{lee77} [L], \citet{stetson00} [S];
{\bf NGC 188} Webda http://obswww.unige.ch/webda/ \citep{webda} [W]; {\bf NGC 288} \citet{alcainoliller80} [A];
{\bf NGC 362} \citet{harris82} [H]; {\bf NGC 1851} \citet{stetson81} [S]; {\bf NGC 1904} \citet{stetsonharris77} [S];
{\bf Berkeley 20} Webda [W]; {\bf NGC 2141} \citet{burkhead72} [B]; \citet{rosvick95} [R]; Webda [W];
{\bf Collinder 110} Webda [W]; {\bf Trumpler 5} Webda [W]; {\bf NGC 2298} \citet{alcainoliller86} [A], \citet{alcaino74} [A];
{\bf Berkeley 32} Webda [W]; {\bf Melotte 66} Webda [W]; {\bf Berkeley 39} Webda [W]; {\bf NGC 2420} Webda [W]; 
{\bf NGC 2506} Webda [W]; {\bf NGC 2682} Webda [W]; {\bf NGC 3201} \citet{stetson00} [S]; 
{\bf NGC 4590} \citet{harris75a} [H], \citet{stetson00} [S]; {\bf NGC 5927} Zoccali (private communication); {\bf NGC 6352} Zoccali (private communication);
{\bf NGC 6528} \citet{ortolani92} [O]; {\bf NGC 6681} \citet{harris75b} [H] \citet{rosenberg00} [R];
{\bf NGC 6705} \citet{sung99} [S]; {\bf NGC 6715} \citet{rosenberg04} [R]; {\bf NGC 6791} \citet{stetsonbg03} [S];
{\bf NGC 6819} Webda [W]; {\bf NGC 7078} \citet{buonanno83} [B]; \citet{stetson00} [S]; {\bf NGC 7789} Webda [W];}
\tablecomments{Table \ref{starsample} is published in its enterety in the electronic edition of Astronomical Journal. A portion is shown here for
guidance regarding its form and content.}

\end{deluxetable}
\clearpage

\begin{deluxetable}{lccccc}
\tabletypesize{\footnotesize}
\tablecaption{Radial velocities of the cluster sample.
\label{samplevr}}
\tablewidth{0pt}
\tablehead{
\colhead{Cluster} & \colhead{V$_r$} & \colhead{$\sigma(V_r)$} &
\colhead{Stars} & \colhead{V$_r$}(ref.) & \colhead{Ref.}}
\startdata
NGC 104 (47 Tuc) & -16 & 11 & 32 & -18.7 & 1 \\
NGC 188 & -44 & 20 &8 & -45 & 2 \\ 	       
NGC 288 & -50 & 11 & 19 & -46.6 & 1 \\	       
NGC 362 & 213 & 7 & 16 & 223.5 & 1 \\	       
NGC 1851 & 321 & 9 & 14 & 320.5 & 1 \\	       
NGC 1904 (M79) & 227 & 5 & 16 & 206 & 1 \\
Berkeley 20 & 80 & 7 & 4 & 70 & 2 \\	       
NGC 2141 & 44 & 10 & 21 & 33/64 & 3,4 \\
Collinder 110 & 45 & 11 & 8 & \nodata & \nodata \\
Trumpler 5 & 44 & 10 & 15 & 54 & 4 \\	   
NGC 2298 & 153 & 15 & 6 & 148.9 & 1 \\	      
Berkeley 32 & 98 & 12 & 3 & 101 & 2 \\	      
Melote 66 & 18 & 10 & 11 & 23 & 5 \\ 	      
Berkeley 39 & 59 & 6 & 5 & 55 & 2 \\	      
NGC 2420 & 69 & 5 & 5 & 67 & 2 \\	
NGC 2506 & 76 & 5 & 3 & 84 & 6 \\	
NGC 2682 (M 67) & 36 & 6 & 9 & 33 & 2 \\
NGC 3201 & 491 & 3 & 10 & 494 & 1 \\ 	      
NGC 4590 (M 68) & -89 & 7 & 19 & -93.4 & 1 \\ 
NGC 5927 & -84 & 5 & 20 & -107.5 & 1 \\	      
NGC 6352 & -114 & 8 & 23 & -121 & 1 \\	      
NGC 6528 & 220 & 7 & 5 & 206 & 1 \\ 	      
NGC 6681 (M 70) & 199 & 7 & 4 & 220 & 1 \\
NGC 6705 (M 11) & 28 & 7 & 10 & 34 & 7 \\
NGC 6715 (M 54) & 156 & 8 & 23 & 142 & 1 \\
NGC 6791 & -46 & 10 & 10 & -57 & 2 \\	      
NGC 6819 & 2 & 5 & 7 & -5 & 2 \\	
NGC 7078 (M 15) & -108 & 10 & 33 & -107 & 1 \\
NGC 7789 & -58 & 6 & 20 & -64 & 2 \\ 	      
\enddata
\tablerefs{
(1) \citet{harris96}; (2) \citet{f02}; (3) \citet{friel89}; (4) \citet{c04} (5) \citet{f93}; (6) \citet{mathieu85};
(7) \citet{mathieu86}}
\end{deluxetable}

\clearpage

\begin{deluxetable}{cc}
\tablecaption{Line and continuum bandpasses
\label{bandastable}}
\tablewidth{0pt}
\tablehead{
\colhead{Line Bandpasses (\AA)} & \colhead{Continuum bandpasses (\AA)}}
\startdata
8484-8513 & 8474-8484\\
8522-8562 & 8563-8577\\
8642-8682 & 8619-8642\\
\nodata & 8799-8725\\
\nodata & 8776-8792\\
\enddata
\end{deluxetable}

\clearpage

\begin{deluxetable}{ccccc}
\tabletypesize{\footnotesize}
\tablewidth{0pt}
\tablecaption{Coefficients of the quadratic fits in the form $\Sigma Ca$=$W'$+$\beta$M$_{V,I}$+$\gamma$M$_{V,I}^2$ to the
sequence of each cluster individually. The clusters are ordered by metallicity.\label{luminositytable}}
\tablehead{
\colhead{Cluster} & \colhead{W'} &\colhead{$\beta$} & \colhead{$\gamma$} & \colhead{$\sigma$}}
\startdata
\multicolumn{5}{c}{M$_V$}\\
NGC 6791 & 10.26$\pm$0.97 & -1.64$\pm$1.28 & 0.38$\pm$0.35 & 0.72\\
NGC 2141 & 8.30$\pm$0.14 & -0.91$\pm$0.13 & 0.02$\pm$0.15 & 0.44\\
NGC 104  & 6.77$\pm$0.06 & -0.69$\pm$0.05 & 0.15$\pm$0.04 & 0.30\\
NGC 288  & 5.46$\pm$0.08 & -0.56$\pm$0.05 & 0.07$\pm$0.03 & 0.29\\
NGC 7078 & 2.79$\pm$0.06 & -0.39$\pm$0.04 & 0.15$\pm$0.04 & 0.22\\
\multicolumn{5}{c}{M$_I$}\\
NGC 6791 & 8.49$\pm$0.24 & -0.63$\pm$0.74 & 0.22$\pm$0.79 & 0.54\\
NGC 2141 & 7.52$\pm$0.11 & -0.73$\pm$0.20 & 0.00$\pm$0.11 & 0.47\\
NGC 104  & 6.30$\pm$0.04 & -0.42$\pm$0.05 & 0.06$\pm$0.02 & 0.14\\
NGC 288  & 5.09$\pm$0.07 & -0.44$\pm$0.04 & 0.03$\pm$0.03 & 0.27\\
NGC 7078 & 2.64$\pm$0.08 & -0.27$\pm$0.14 & 0.05$\pm$0.07 & 0.22\\
\tableline
\enddata
\end{deluxetable}

\clearpage

\begin{deluxetable}{lcccc}
\tabletypesize{\footnotesize}
\tablewidth{0pt}
\tablecaption{Derived $W'_V$ and $W'_I$ and number of stars used.\label{fitstable}}
\tablehead{
\colhead{Cluster} & \colhead{$W'_V$} & \colhead{\#Star} & \colhead{$W'_I$} &
\colhead{\#Star}}
\startdata
NGC 104 & $6.94\pm0.01 $ & 34 & $6.23\pm0.01$ & 14\\
NGC 188 & $8.17\pm0.07 $ & 6 & $7.27\pm0.08$ & 5\\
NGC 288 & $5.51\pm0.01 $ & 19 & $5.04\pm0.03$ & 14\\
NGC 362 & $6.01\pm0.01 $ & 16 & \nodata & \nodata\\
NGC 1851 & $5.94\pm0.03 $ & 14 & $5.24\pm0.04$ & 8\\
NGC 1904 & $4.91\pm0.03 $ & 16 & \nodata & \nodata\\
Berkeley 20 & $6.86\pm0.03 $ & 4 & $6.39\pm0.03$ & 3\\
NGC 2141 & $8.33\pm0.01 $ & 18 & $7.67\pm0.02$ & 15\\
Collinder 110 & $8.21\pm0.04 $ & 11 & $7.74\pm0.06$ & 6\\
Trumpler 5 & $7.52\pm0.04 $ & 16 & $6.97\pm0.04$ & 15\\
NGC 2298 & $3.75\pm0.03 $ & 6 & $3.09\pm0.03$ & 5\\
Berkeley 32 & $5.90\pm0.08 $ & 4 & $5.27\pm0.08$ & 4\\
Melote 66 & $7.69\pm0.03 $ & 11 & $6.90\pm0.03$ & 11\\
Berkeley 39 &$8.21\pm0.04 $ & 5 & $7.27\pm0.06$ & 3\\
NGC 2420 & $ 6.26\pm0.09 $ & 6 & $6.15\pm0.08$ & 4\\
NGC 2506 & $ 6.96\pm0.09 $ & 4 & $6.37\pm0.09$ & 3\\
NGC 2682 & $ 8.24\pm0.01 $ & 6 & $7.48\pm0.01$ & 8\\
NGC 3201 & $ 5.46\pm0.03 $ & 9 & $4.76\pm0.02$ & 6 \\
NGC 4590 & $ 2.84\pm0.02 $ & 19 & $2.19\pm0.06$ & 12\\
NGC 5927 & $ 7.81\pm0.01 $ & 21 & $6.92\pm0.01$ & 13\\
NGC 6352 & $ 7.10\pm0.01 $ & 19 & $6.31\pm0.01$ & 19\\
NGC 6528 & $ 8.28\pm0.04 $ & 5 & $7.58\pm0.04$ & 5\\
NGC 6681 & $ 5.05\pm0.03 $ & 4 & $4.49\pm0.07$ & 3\\
NGC 6705 & $ 8.95\pm0.07 $ & 7 & $8.28\pm0.12$ & 6\\
NGC 6715 & $ 4.86\pm0.03 $ & 23 & $4.30\pm0.03$ & 24\\
NGC 6791 & $ 9.78\pm0.09 $ & 9 & $8.77\pm0.09$ & 8\\
NGC 6819 & $ 8.41\pm0.04 $ & 7 & $7.64\pm0.04$ & 7\\
NGC 7078 & $ 2.78\pm0.01 $ & 38 & $2.18\pm0.01$ & 14\\
NGC 7789 & $ 8.47\pm0.02 $ & 20 & $7.61\pm0.02$ & 20\\
\enddata
\end{deluxetable}

\end{document}